\documentclass[a4paper,fleqn,usenatbib]{mnras}
\usepackage{newtxtext,newtxmath}
\usepackage[T1]{fontenc}
\usepackage{ae,aecompl}
\usepackage{graphicx}	
\usepackage{amsmath}	
\usepackage{amssymb}	
\usepackage{ulem}
\usepackage{ascmac}
\usepackage{bm}

\newcommand{\Msun}{$\rm M_{\odot}$}
\newcommand{\mvir}{$M_{\rm vir}$}
\newcommand{\pc}{$\rm pc$}
\newcommand{\Afs}{A\_N8192L800}
\newcommand{\Af}{A\_N4096L400}
\newcommand{\As}{A\_N4096L200}
\newcommand{\Bfs}{B\_N4096L400}
\newcommand{\Bf}{B\_N2048L200}
\newcommand{\nugc}{$\nu^2$GC}

\title[Subhaloes near the Free Streaming Scale]
{The Abundance and Structure of Subhaloes near the Free Streaming Scale 
and Their Impact on Indirect Dark Matter Searches}

\author[Ishiyama et al.]{
Tomoaki Ishiyama$^{1}$\thanks{E-mail: ishiyama@chiba-u.jp}
and Shin'ichiro Ando$^{2,3}$
\\
$^{1}$Institute of Management and Information Technologies, Chiba University, 1-33, Yayoi-cho, Inage-ku, Chiba, 263-8522, Japan \\
$^{2}$GRAPPA Institute, University of Amsterdam, 1098 XH Amsterdam, Netherlands \\
$^{3}$Kavli Institute for the Physics and Mathematics of the Universe (Kavli IPMU, WPI), \\
Todai Institutes for Advanced Study, University of Tokyo, Kashiwa, Chiba 277-8583, Japan
}

\date{Accepted XXX. Received YYY; in original form ZZZ}
\pubyear{XXX}

\begin{document}
\label{firstpage}
\pagerange{\pageref{firstpage}--\pageref{lastpage}}
\maketitle 

\begin{abstract} 
The free streaming motion of dark matter particles 
imprints a cutoff in the matter power spectrum and set the scale of the 
smallest dark matter halo. Recent cosmological $N$-body simulations have shown that 
the central density cusp is much steeper in haloes near the free 
streaming scale than in more massive haloes.  Here, we study the 
abundance and structure of subhaloes near the free streaming scale 
at very high redshift
using a suite of unprecedentedly large cosmological $N$-body simulations, over 
a wide range of the host halo mass.  The subhalo abundance is suppressed 
strongly below the free streaming scale, but the ratio between the 
subhalo mass function in the cutoff and no cutoff simulations is well 
fitted by a single correction function regardless of the host halo mass 
and the redshift.  In subhaloes, the central slopes are considerably 
shallower than in field haloes, however, are still steeper than that of 
the NFW profile.  Contrary, the concentrations are significantly larger 
in subhaloes than haloes and depend on the subhalo mass. 
We compare two 
methods to extrapolate the mass-concentration relation of haloes and 
subhaloes to z=0 and provide a new simple fitting function for subhaloes, 
based on a suite of large cosmological $N$-body simulations.
Finally, we estimate the annihilation boost factor of a Milky-Way sized halo
to be between 1.8 and 6.2.
\end{abstract}

\begin{keywords}
cosmology: theory
---methods: numerical
---Galaxy: structure
---dark matter
\end{keywords}

\section{Introduction}\label{sec:intro}

The smallest dark matter haloes are the first gravitationally collapsed 
structures in the Universe according to the hierarchical structure 
formation scenario.  The free streaming motion of particles imprints a 
cutoff in the matter power spectrum at the initial stage of the 
Universe and determines the size of the smallest haloes. If dark matter 
consists of 
Weakly Interacting Massive Particle (WIMP) of mass approximately 100 Ge,V
their mass has been estimated to be
near Earth-mass, $10^{-12}\mbox{\scriptsize --}10^{-3}$\Msun\ 
\citep[e.g.,][]{Zybin1999, Hofmann2001, Berezinsky2003,
  Green2004, Loeb2005, Bertschinger2006, Profumo2006, Berezinsky2008,
  Diemanti2015}. 

In this decade, the density profiles of haloes near the free streaming
scale have been studied by means of cosmological $N$-body simulations
\citep{Diemand2005, Ishiyama2010, Anderhalden2013, Ishiyama2014,
  Angulo2017, Delos2019}, merger simulations \citep{Ogiya2016,Angulo2017},
cold collapse simulations \citep{Ogiya2018},
and idealized tidal evolution simulations \citep{Delos2019b}. 
\citet{Ishiyama2010}
found that the central density cusps are considerably steeper in
microhaloes than more massive haloes and are well described by
$\rho(r)\propto r^{-1.5}$.  These results are supported by other
similar cosmological simulations using different simulation
codes \citep{Anderhalden2013, Angulo2017} and are reproduced
by cold collapse simulations \citep{Ogiya2018}.
\citet{Ishiyama2014} (hereafter I14) extended these results with
better statistics and found that the cusp slope gradually becomes
shallower with increasing halo mass, as a result of major merger
processes \citep[see also][]{Ogiya2016,Angulo2017}.  Similar profiles
are obtained in warm dark matter simulations \citep{Polisensky2015},
in which the cutoff in the matter power spectrum is also imposed
although its corresponding mass scale is much larger than that of
microhaloes. Steeper cusps are also observed in recent simulations of
ultracompact minihaloes \citep{Gosenca2017, Delos2018b, Delos2018}.

Such steep cusps may have a significant effect on dark matter searches.  
There are a variety of subjects such as
gravitational lensing \citep{Chen2010, Erickcek2011, Tilburg2018},
Gravitational Waves \citep{Bird2016}, Galactic tidal fluctuations
\citep{Penarrubia2018}, gravitational perturbations on the Solar
system \citep{Gonzalez2013}, pulsar timing arrays \citep{Ishiyama2010,
  Baghram2011, Kashiyama2018} and indirect detection experiments
 (e.g., \citealt{Berezinsky2003, Koushiappas2004, Koushiappas2006,
  Goerdt2007, Diemand2007, Ando2008, Ishiyama2014, Bartels2015,
  Fornasa2015, Anderson2016, Marchegiani2016, Hutten2016, Hutten2018,
  Hooper2017, Moline2017, Stref2017, Hiroshima2018, Kamada2019, Karwin2019}, 
and see also a recent review by \citealt{Ando2019}). 
I14 showed that 
the steeper inner cusps of haloes near the free streaming scale can 
increase the dark matter annihilation luminosity of a Milky-Way sized 
halo between 12 \% to 67 \%, compared with the case we assume the NFW 
density profile \citep{Navarro1997} and an empirical mass-concentration 
relation proposed by \citet{Sanchez-Conde2014}.  However, this 
estimation relies on density profiles seen in field haloes, not subhaloes. 
 To make a more robust estimation, quantifying the structures of subhaloes
near the free streaming scale is necessary.

Not the density structure but the abundance of microhaloes in the
Milky-Way halo are crucial for the annihilation signal.  Analytic
studies and cosmological simulations have suggested that the subhalo
mass function 
is expressed as $dn/dm \propto m^\zeta$, 
with the slope $\zeta$ of $-(2\mbox{\scriptsize --}1.8)$
\citep[e.g.,][]{Hiroshima2018}, although
there is no consensus.  The cutoff in the matter power spectrum should
suppress the number of subhaloes near the free streaming scale, which
should weaken the annihilation signal.  However, the shape of the mass
function near the free streaming scale is not understood well, 
in particular, for the case of WIMP dark matter.

We address these questions by large and high resolution cosmological
$N$-body simulations.  This paper presents the first challenge to reveal the
structure of subhaloes near the free streaming scale. To reliably
study the statistics of these subhaloes, we
conducted huge simulations with sufficient spatial volumes.  
\S~\ref{sec:method} describes our simulation method and its setup.  The
mass function, density profiles, and concentrations of subhaloes
near the free streaming are presented in \S~\ref{sec:result}. The
contributions of these haloes to gamma-ray signals by WIMP
self-annihilation is discussed in \S~\ref{sec:discussion}.  The
results are summarized in \S~\ref{sec:summary}.

\section{Initial Conditions and Numerical Method}\label{sec:method}

\begin{table*}
\centering
\caption
{Parameters of a suite of cosmological $N$-body simulations 
performed in this work. Those of three simulations conducted in I14 
are also shown for the comparison. 
 Here, $N$, $L$, $\varepsilon$, $M_{\rm total}$, 
 and $m_{\rm p}$
are the total number of particles, box length, softening length, total mass in the box, 
and particle mass resolution, respectively.
The final redshift when the simulations are terminate is $z=32$. }
\label{tab1}
\begin{tabular}{lccccccc}
\hline
Name  & $N$ & $L (\pc)$ & $\varepsilon (\pc)$ & $M_{\rm total}$ (\Msun) & $m_{\rm p}$ (\Msun) & cutoff & reference\\
\hline
\Afs & $8192^3$ & 800.0 & $2.0 \times 10^{-4}$ & 20.37 & $3.7 \times 10^{-11}$ & w/ & this work \\
\Af & $4096^3$ & 400.0 & $2.0 \times 10^{-4}$ & 2.35 & $3.4 \times 10^{-11}$ & w/ & \citet{Ishiyama2014}\\
\As & $4096^3$ & 200.0 & $1.0 \times 10^{-4}$ & 0.30 & $4.3 \times 10^{-12}$ & w/ & \citet{Ishiyama2014}\\
\Bfs & $4096^3$ & 400.0 & $2.0 \times 10^{-4}$ & 2.55 & $3.7 \times 10^{-11}$ & w/o & this work\\
\Bf & $2048^3$ & 200.0 & $2.0 \times 10^{-4}$ & 0.30 & $3.4 \times 10^{-11}$ & w/o & \citet{Ishiyama2014}\\
\hline 
\end{tabular}
\end{table*}

\begin{table*}
\centering
\caption
{Same as Table~\ref{tab1}. Here, $z_{\rm fin}$ is the redshift when the simulation is terminated. These five simulations do not account the cutoff in the matter power spectrum.}
\label{tab2}
\begin{tabular}{lccccccc}
\hline
Name  & $N$ & $L (h^{-1} \rm Mpc)$ & $\varepsilon (h^{-1} \rm kpc)$ & $M_{\rm total} (h^{-1} \rm M_{\odot})$ & $m_{\rm p} (h^{-1} \rm M_{\odot})$ & $z_{\rm fin}$ & reference \\
\hline 
\nugc-S & $2048^3$ & 280.0 & $4.28$ & $1.89 \times 10^{18}$ & $2.20 \times 10^{8}$ & 0.0 & \citet{Ishiyama2015}\\
\nugc-H2 & $2048^3$ & 70.0 & $1.07$ & $2.95 \times 10^{16}$ & $3.44 \times 10^{6}$ & 0.0 & \citet{Ishiyama2015}\\
Phi-0 & $2048^3$ & 8.0 & $0.12$ & $4.40 \times 10^{13}$ & $5.13 \times 10^{3}$ & 0.0 & \citet{Ishiyama2016}\\
Phi-1 & $2048^3$ & 32.0 & $0.48$ & $2.82 \times 10^{15}$ & $3.28 \times 10^{5}$ & 0.0 & this work \\
Phi-2 & $2048^3$ & 1.0 & $7.5 \times 10^{-3}$ & $8.60 \times 10^{10}$ & $10.0$ & 5.0 & this work\\
\hline 
\end{tabular}
\end{table*}

To follow the formation and evolution of haloes and subhaloes near the 
free streaming scale, we conducted two ultralarge cosmological $N$-body 
simulations. The matter power spectrum in the simulation denoted \Afs\ 
accounted the cutoff imposed by the free motion of dark matter particles
\citep{Green2004}.  
The cutoff scale is nearly earth-mass, $10^{-6}$\Msun.
The other simulation 
denoted \Bfs\ ignored the effect of the free motion. The cosmological 
parameters are $\Omega_0=0.31$, $\rm \lambda_0=0.69$, $h=0.68$, $n_{\rm 
s}=0.96$, and $\sigma_8=0.83$, which are consistent with an observation 
of the cosmic microwave background obtained by the Planck satellite 
\citep{Planck2014,Planck2016,Planck2018}. The initial conditions were generated by 
a first-order Zeldovich approximation at $z=400$. Note that the power 
spectrum near the cutoff scale can be increased in early matter 
dominated era \citep{Erickcek2015}, which we do not consider here.

In the simulations, \Afs\ and \Bfs, 
the gravitational evolution of $8192^3$ and $4096^3$ particles 
in comoving boxes of side length 800 pc and 400 pc 
were calculated, respectively. 
The particle masses of both simulations are
$3.7 \times 10^{-11} M_{\odot}$, comparable to those of I14.
The gravitational softening length is $2.0 \times 10^{-4}$ pc. 
The parameters of the two simulations and our earlier ones \citep{Ishiyama2014} 
are listed in Table~\ref{tab1}.

The gravitational evolution was followed by a massively parallel
TreePM code, GreeM \citep{Ishiyama2009b,
  Ishiyama2012}\footnote{http://hpc.imit.chiba-u.jp/\~{}ishiymtm/greem/}
on the K computer at the RIKEN Advanced Institute for Computational
Science, Aterui and Aterui II supercomputer at Center for Computational
Astrophysics, CfCA, of National Astronomical Observatory of Japan.
The evaluation of the tree force was accelerated by the Phantom-GRAPE
software\footnote{http://code.google.com/p/phantom-grape/}
\citep{Nitadori2006, Tanikawa2012, Tanikawa2013, Yoshikawa2018} 
with support for the
HPC-ACE architecture of the K computer.  These simulations were
terminated at $z=32$.

Haloes were identified by the spherical overdensity method
\citep{Lacey1994}.  To calculate the virial mass of a halo \mvir, we
used the overdensity, $\Delta(z)=18\pi^2+ 82x - 39x^2$ times the
critical density in the Universe, where $x\equiv \Omega(z)-1$, based
on the spherical collapse model \citep{Bryan1998}.  The most massive
haloes identified in \Afs\ and \Bfs\ simulations contain 796,727,804
and 504,029,713 particles, corresponding $2.95 \times
10^{-2}$\Msun\ and $1.87 \times 10^{-2}$\Msun, respectively.
Subhaloes were identified by ROCKSTAR phase space halo/subhalo finder
\footnote{https://bitbucket.org/gfcstanford/rockstar/}
\citep{Behroozi2013}.  The total number of haloes and subhaloes with
masses larger than $10^{-6}$\Msun\ are 32,907 and 22,936 for the
\Afs, and 20,915 and 3,987 for the \Bfs\ at $z=32$.

To compare the structures of haloes and subhaloes near the free
streaming scale with those of more massive haloes, we used three
simulations taken from \citet{Ishiyama2015, Ishiyama2016, Makiya2016},
and conducted two new simulations. These five simulations do not
account the cutoff in the matter power spectrum.  The parameters of
these simulations are summarized in Table~\ref{tab2}.

To generate the initial conditions for the $\nu^2$GC-S, $\nu^2$GC-H2, and 
Phi-0 simulations, we used a publicly available code,
2LPTic\footnote{http://cosmo.nyu.edu/roman/2LPT/}.
The initial conditions for the Phi-1 and Phi-2 simulations 
were generated by a publicly available code, MUSIC
\footnote{https://bitbucket.org/ohahn/music/}\citep{Hahn2011}. 
Both codes adopt second-order Lagrangian
perturbation theory \citep[e.g.,][]{Crocce2006}. 
The other numerical method and cosmological parameters used in these 
simulations are the same as described above.
The Rockstar catalogues and consistent merger trees of the
$\nu^2$GC-S, $\nu^2$GC-H2, and Phi-1 simulations are publicly available at
\url{http://hpc.imit.chiba-u.jp/~ishiymtm/db.html}.

\section{Results}\label{sec:result}

\subsection{Subhalo Mass Function}\label{sec:massfunc}

\begin{figure*}
\centering 
\includegraphics[width=8.8cm]{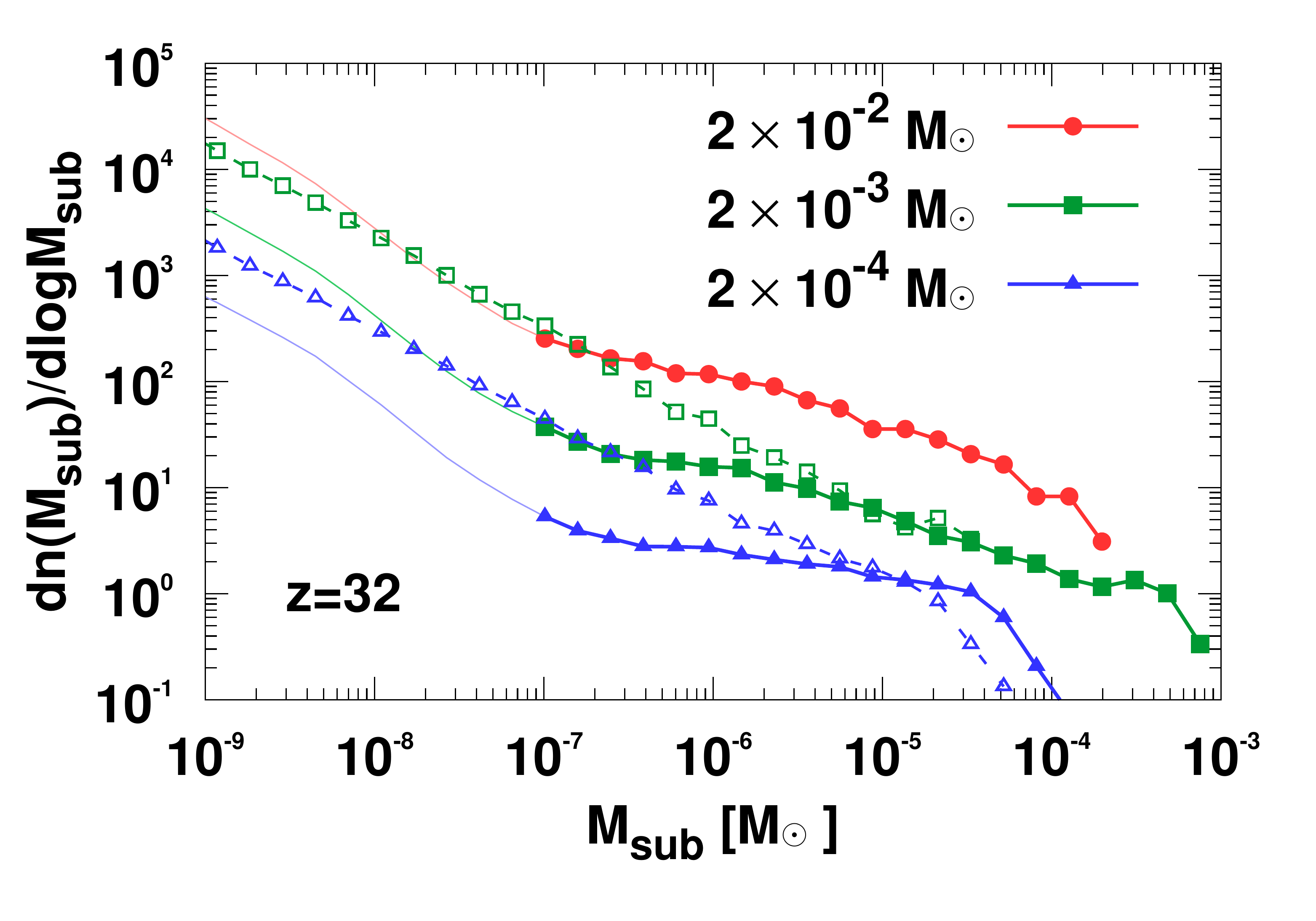} 
\includegraphics[width=8.8cm]{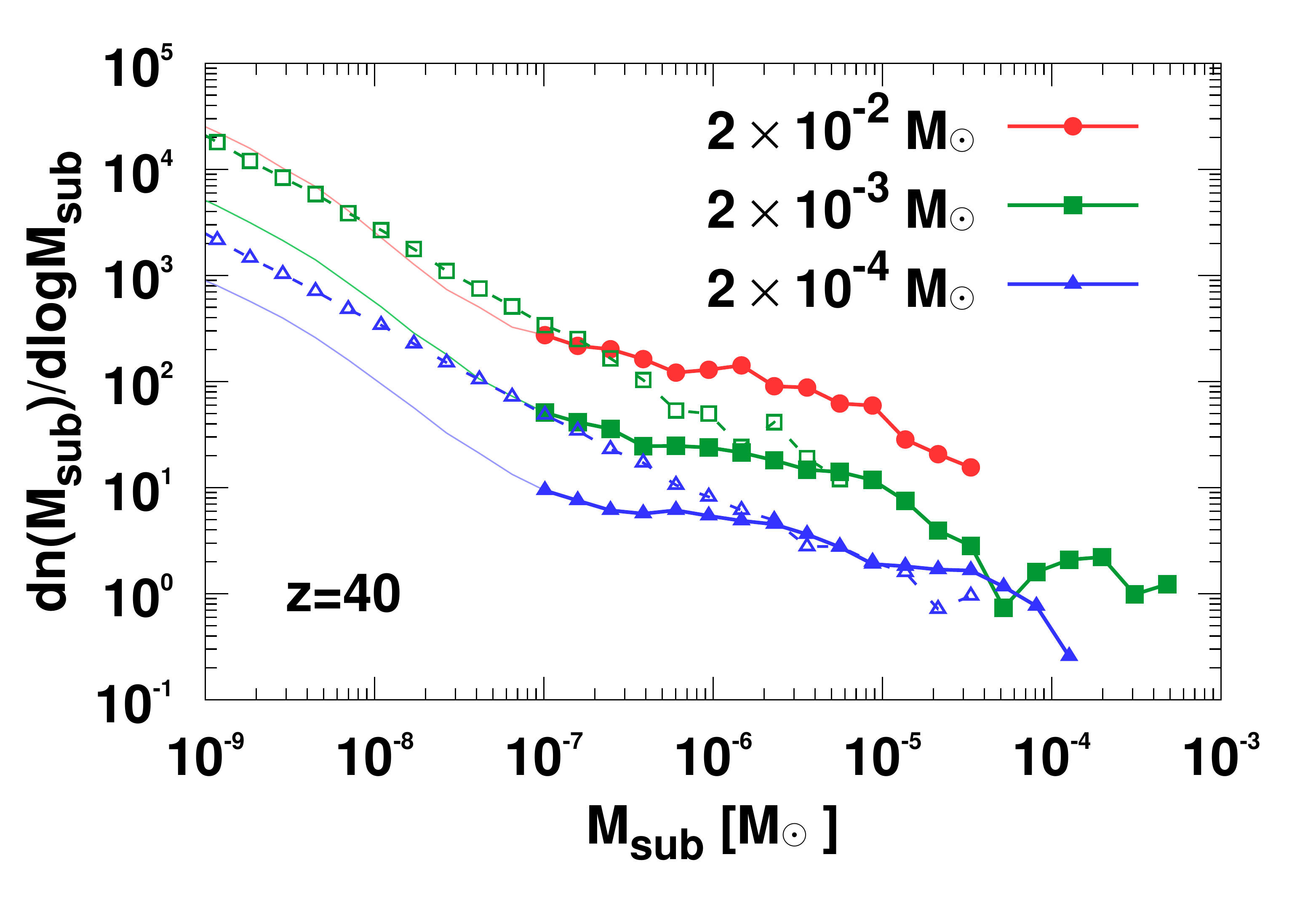} 
\caption{ Stacked subhalo mass functions of haloes for three different
  mass bins, $2 \times 10^{-2}$\Msun, $2 \times 10^{-3}$\Msun, and $2
  \times 10^{-4}$\Msun, at $z=32$ (left) and 40 (right).
  Solid curves show the result of the simulation with the cutoff, \Afs.  Dashed curves show the
  result of the simulation without the cutoff, \Bfs.  
For subhaloes less massive than $\sim 5.0 \times 10^{-6}$\Msun\ in 
the \Afs\ simulation, 
the slopes of mass functions $\zeta$ are gradually
decreasing with decreasing subhalo mass differently from the
\Bfs\ simulation and are becoming flat at around $\sim 10^{-6}$\Msun,
which corresponds to the cutoff scale.  The slopes $\zeta$ are
becoming steeper again below $\sim 10^{-7}$\Msun\ 
(displayed by thin curves) due to artificial
fragmentation as observed in simulations of warm dark matter \citep{Wang2007,
  Schneider2013, Angulo2013}.
}
\label{fig:subhalo_massfunc}
\end{figure*}

We selected subhaloes found within the virial radius of their host haloes 
and calculated the subhalo mass function of each halo.  To derive the 
proper average mass functions of haloes with vast mass distribution, we 
stacked the mass function of similar-mass haloes. 
Figure~\ref{fig:subhalo_massfunc} shows the stacked subhalo mass 
functions of the simulations \Afs\ and \Bfs\ for three different ranges 
of the host halo mass, $1.00\mbox{--}3.16 \times 
10^{-2}$\Msun, $1.00\mbox{--}3.16 \times 10^{-3}$\Msun, and 
$1.00\mbox{--}3.16 \times 10^{-4}$\Msun\ at $z=32$ and 40.  For the 
\Bfs\ simulation, the stacked subhalo mass function of halo with the 
mass $1.00\mbox{--}3.16 \times 10^{-2}$\Msun\ is not displayed because of the 
absence of the host haloes.

The mass functions without the cutoff (\Bfs) show a nearly single
power law except for the massive end, consistent with more massive host
haloes and lower redshifts \citep[e.g.,][]{Hiroshima2018}.  The mass
functions with the cutoff (\Afs) agree with those of the \Bfs\ for the
subhalo mass more massive than $\sim 5.0 \times 10^{-6}$\Msun.  For
the less massive subhaloes, the slopes of mass functions $\zeta$ are gradually
decreasing with decreasing subhalo mass differently from the
\Bfs\ simulation and are becoming flat at around $\sim 10^{-6}$\Msun,
which corresponds to the cutoff scale.  Interestingly, the slopes $\zeta$ are
becoming steeper again below $\sim 10^{-7}$\Msun\ due to artificial
fragmentation as observed in simulations of warm dark matter 
\citep{Wang2007, Schneider2013, Angulo2013}.

\begin{figure}
\centering 
\includegraphics[width=9cm]{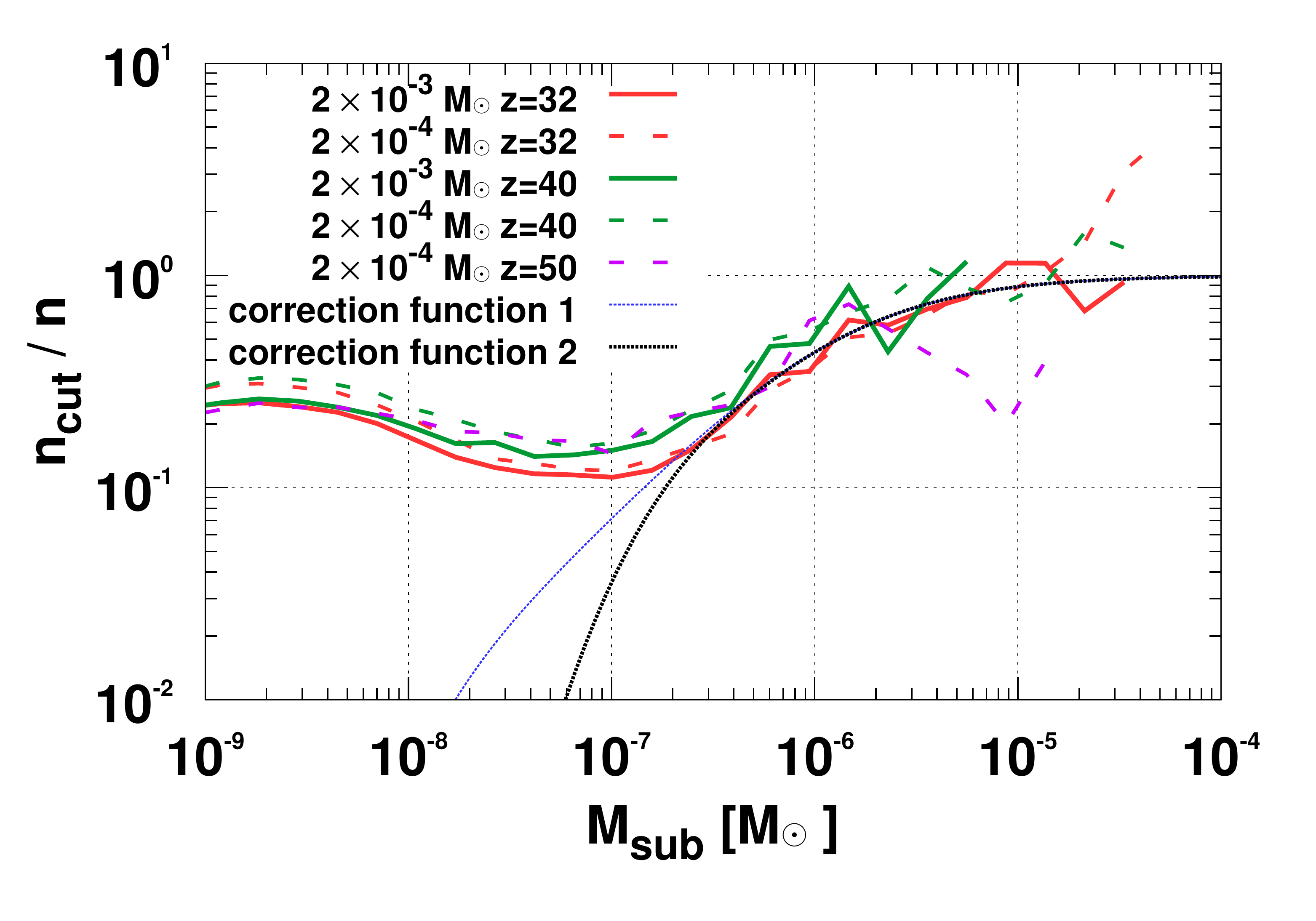} 
\caption{ 
Effect of the cutoff in the matter power spectrum 
on the subhalo mass functions at $z=32$, 40, and 50: 
the ratio between the mass function with (\Afs) and without (\Bfs) 
the cutoff. 
Thick and thin dotted curves show the two different correction functions
to convert the mass function without the cutoff into with the cutoff 
as described in Equation~(\ref{eq:fcor}).
}
\label{fig:subhalo_massfunc_comp}
\end{figure}

These results are highlighted in Figure~\ref{fig:subhalo_massfunc_comp}, 
which shows the ratio between the subhalo mass
function with (\Afs) and without (\Bfs) the cutoff at $z=32$, 40, and 50.  
As shown in Figure~\ref{fig:subhalo_massfunc}, the ratios are decreasing and
flattening with decreasing subhalo mass, and show upturns at around
$\sim 10^{-7}$\Msun\ due to artificial fragmentation.  The shapes of
ratios agree well with each other regardless of the host halo mass and the
redshift. 

The ratio is well fitted by the correction function, 
\begin{eqnarray}
{\rm f_{cor}}(M) = \frac{n_{\rm cut}}{n} = \frac{1}{2} \left( 1+\frac{M_1}{M} \right)^{-1} \left[1 + {\rm erf}\left( \ln\frac{M}{M_2}\right) \right], 
\label{eq:fcor}
\end{eqnarray}
where, $M_1=1.3 \times 10^{-6}$, and $M_2=1.0 \times 10^{-8}$, or $1.0
\times 10^{-7}$\Msun.
Hereafter, we call the former "correction function 1" and 
the later "correction function 2".
To remove the influence of
artificial fragmentation, dumping in the correction function is
necessary around the free streaming scale.  Because it is difficult to
distinguish physical haloes from artificial fragmentation, we introduce
two correction functions with different $M_2$ and compare the effect
on the annihilation signal in \S~\ref{sec:discussion}.
From the nature of free streaming damping, 
this correction function should be valid for any mass scales
more massive than $10^{-8}$\Msun\ for the correction function 1 
and $10^{-7}$\Msun\ for the correction function 2. 

This functional form is the same introduced in \citet{Angulo2013}, which 
used it to fit the ratio of halo mass function in warm dark matter to 
cold dark matter simulations.  This similarity seems to be a natural 
consequence because the physical origin of the dumping in the matter 
power spectrum is the same in both our simulations and warm dark matter 
simulations although there are more than ten orders of difference 
between their free streaming scales. 

From the independence of the subhalo mass function near the free
streaming scale on the host halo mass and the redshift, the assumption
should be justified that the shape of subhalo mass functions for more
massive host haloes and lower redshifts should be similar to what we
see in Figure~\ref{fig:subhalo_massfunc}.  By extrapolating a
subhalo mass function with a power law down to the smallest scale and
multiplying the ratio shown in Figure~\ref{fig:subhalo_massfunc_comp}
to it, 
we should be able to predict the mass function 
at an arbitrary redshift and of host halo mass, 
from the smallest to the largest scale
under some assumptions for its normalization. 
This assumption is supported by analytic models
\citep[e.g.,][]{Hiroshima2018} that have found that the power law index of the
subhalo mass function is in a rather narrow range between $-2$ and $-1.8$ with a
vast range of halo/subhalo mass from $z=0$ to 5.

\subsection{Subhalo Density Profiles}\label{sec:profile}

I14 have shown that the central density cusps are
substantially steeper in haloes near the free streaming scale
than more massive haloes when the cutoff is imposed in the matter power
spectrum.  
A double power law function, given by
\begin{eqnarray}
\rho(r) = \frac{\rm \rho_0}{ (r/r_{\rm s})^{\alpha} (1+r/r_{\rm s})^{(3-\alpha)}},
\label{eq:doublepower}
\end{eqnarray} 
fits density profiles better than NFW and Einasto profiles.  The
dependence of $\alpha$ and concentration parameter $c=r_{\rm
  vir}/r_{\rm s}$ on the halo mass is given in I14.  The cusp slope
gradually becomes shallower with increasing halo mass through major
merger processes \citep[see also][]{Ogiya2016, Angulo2017}.
The concentration parameter slightly depends on the halo mass.

\begin{figure*}
\centering 
\includegraphics[width=8.8cm]{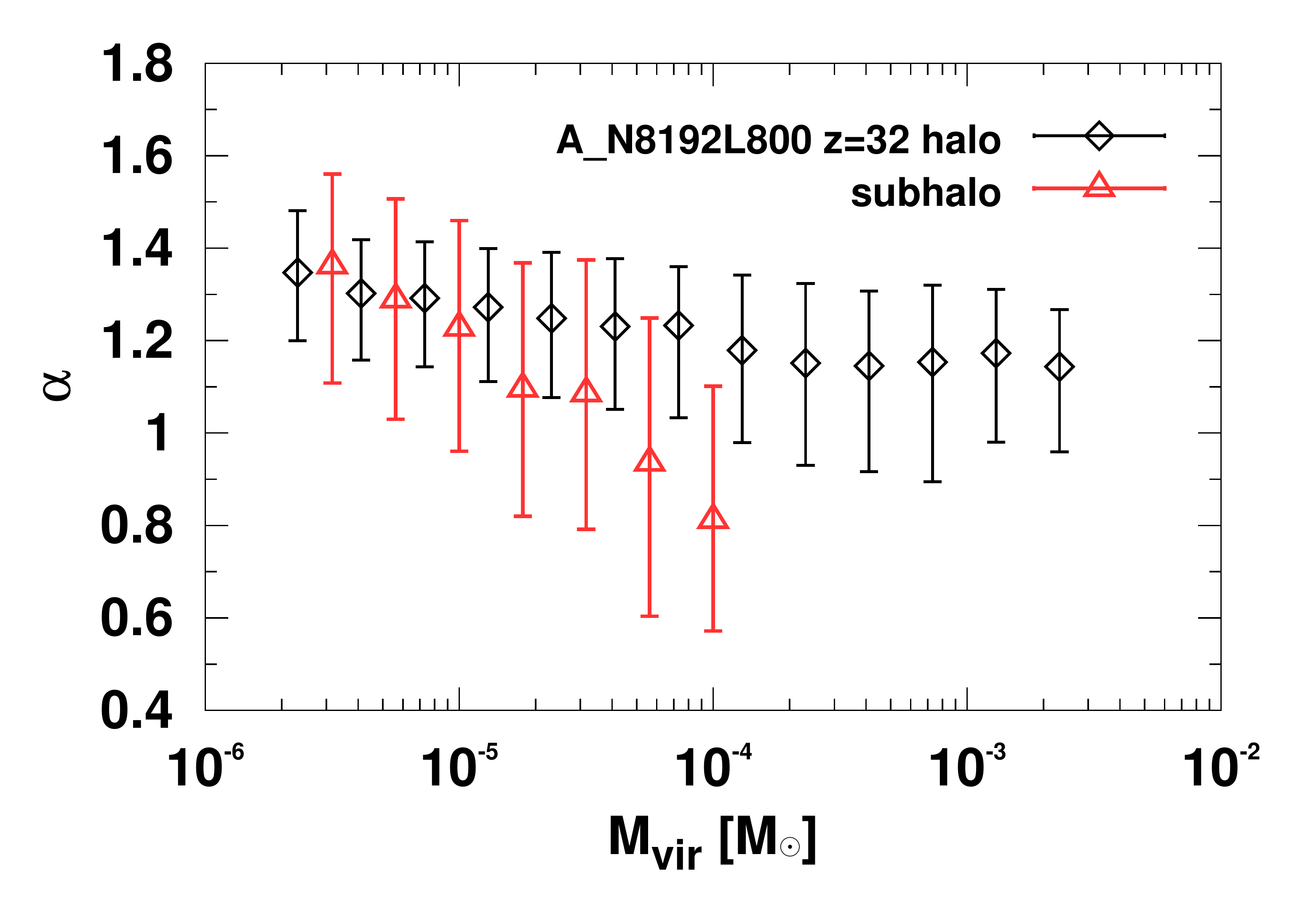} 
\includegraphics[width=8.8cm]{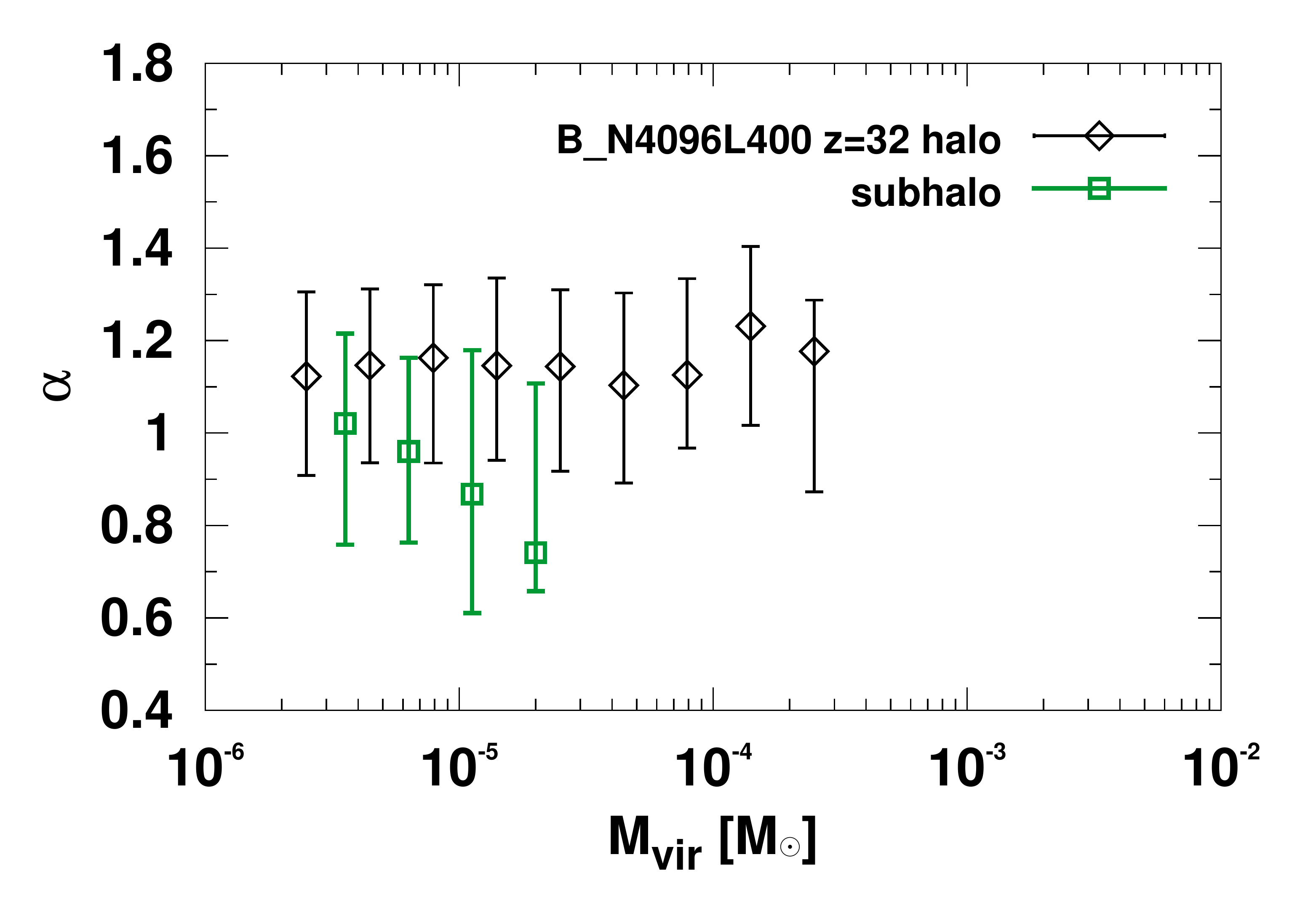} 
\caption{ 
Slopes of the density profiles of field haloes and subhaloes $\alpha$ as a function of 
the virial mass \mvir\ at $z=32$, 
for the \Afs\ (left) and \Bfs\ (right) simulations.
Triangles, squares, and crosses are the median value in each mass bin.
Whiskers give the first and third quantiles. 
  }
\label{fig:profile_a}
\end{figure*}

\begin{figure*}
\centering 
\includegraphics[width=8.5cm]{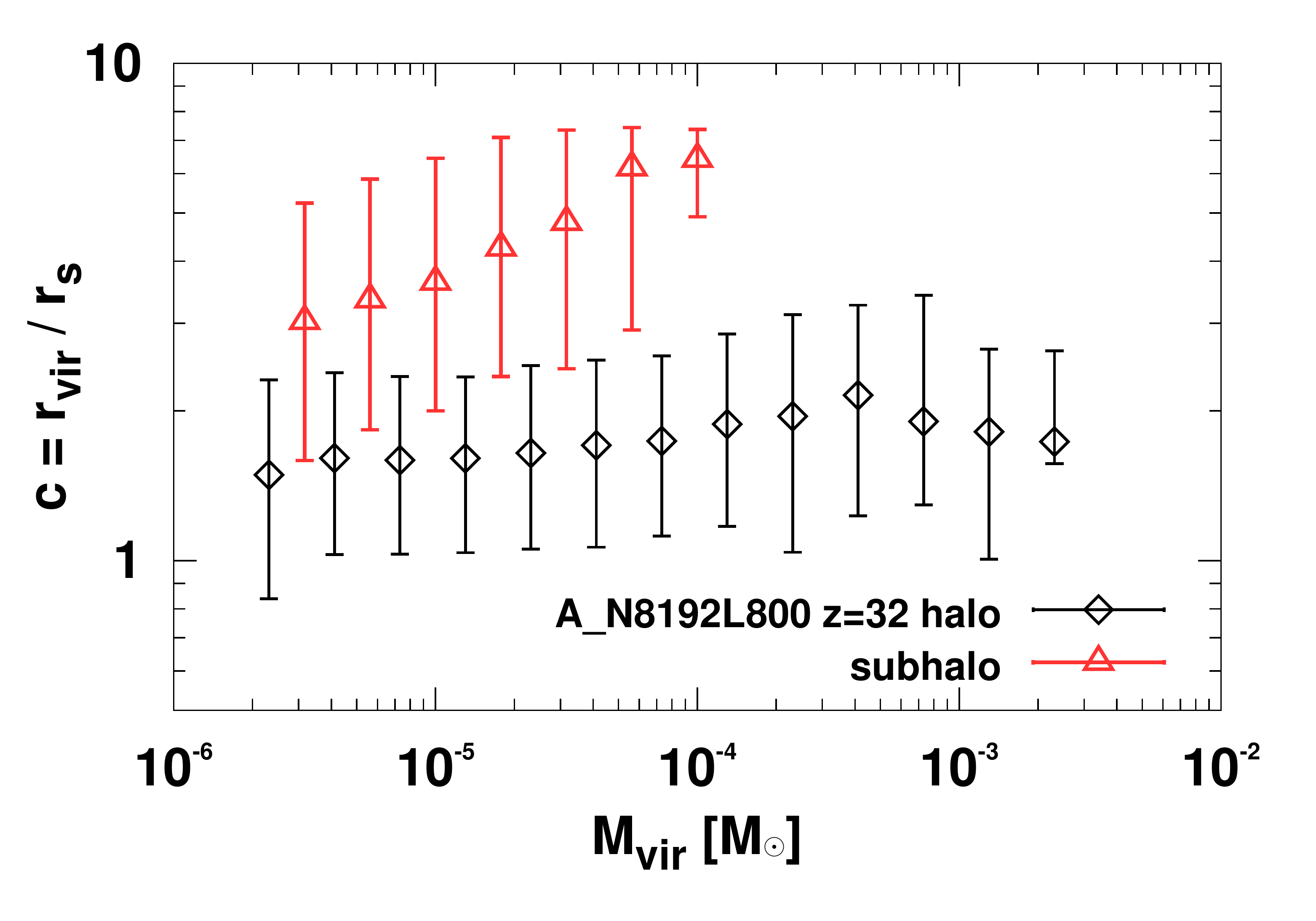}
\includegraphics[width=8.5cm]{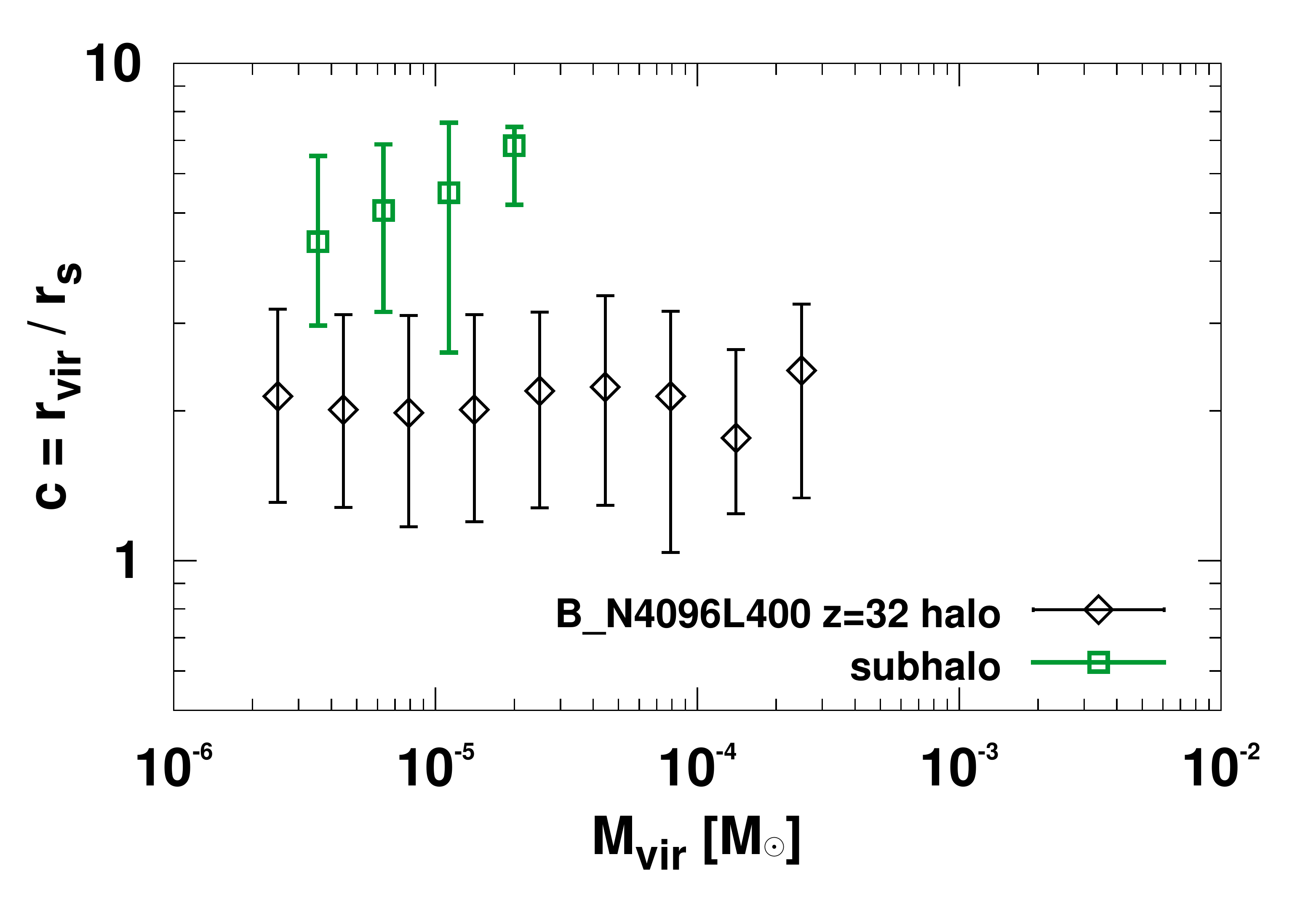}
\caption{ 
Concentrations of the density profiles of field haloes and subhaloes
  $r_{\rm vir}/r_{\rm s}$ as a function of the virial mass \mvir\ at
  $z=32$, for the \Afs\ (left) and \Bfs\ (right) simulations.
  Triangles, squares, and crosses 
  are the median value in each mass bin.  Whiskers give
  the first and third quantiles. 
}
\label{fig:profile_c}
\end{figure*}

\begin{figure*}
\centering 
\includegraphics[width=8.5cm]{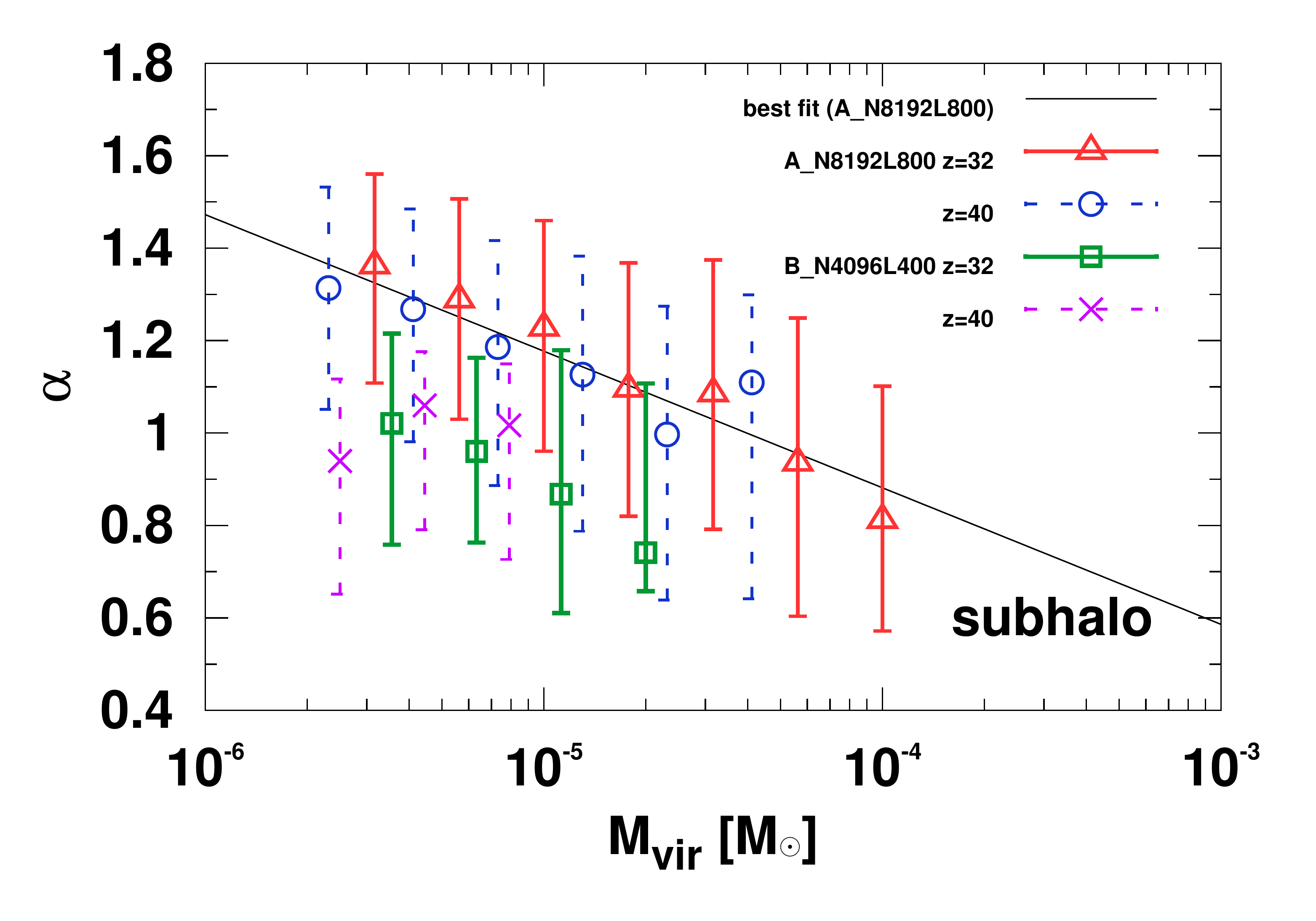}
\includegraphics[width=8.5cm]{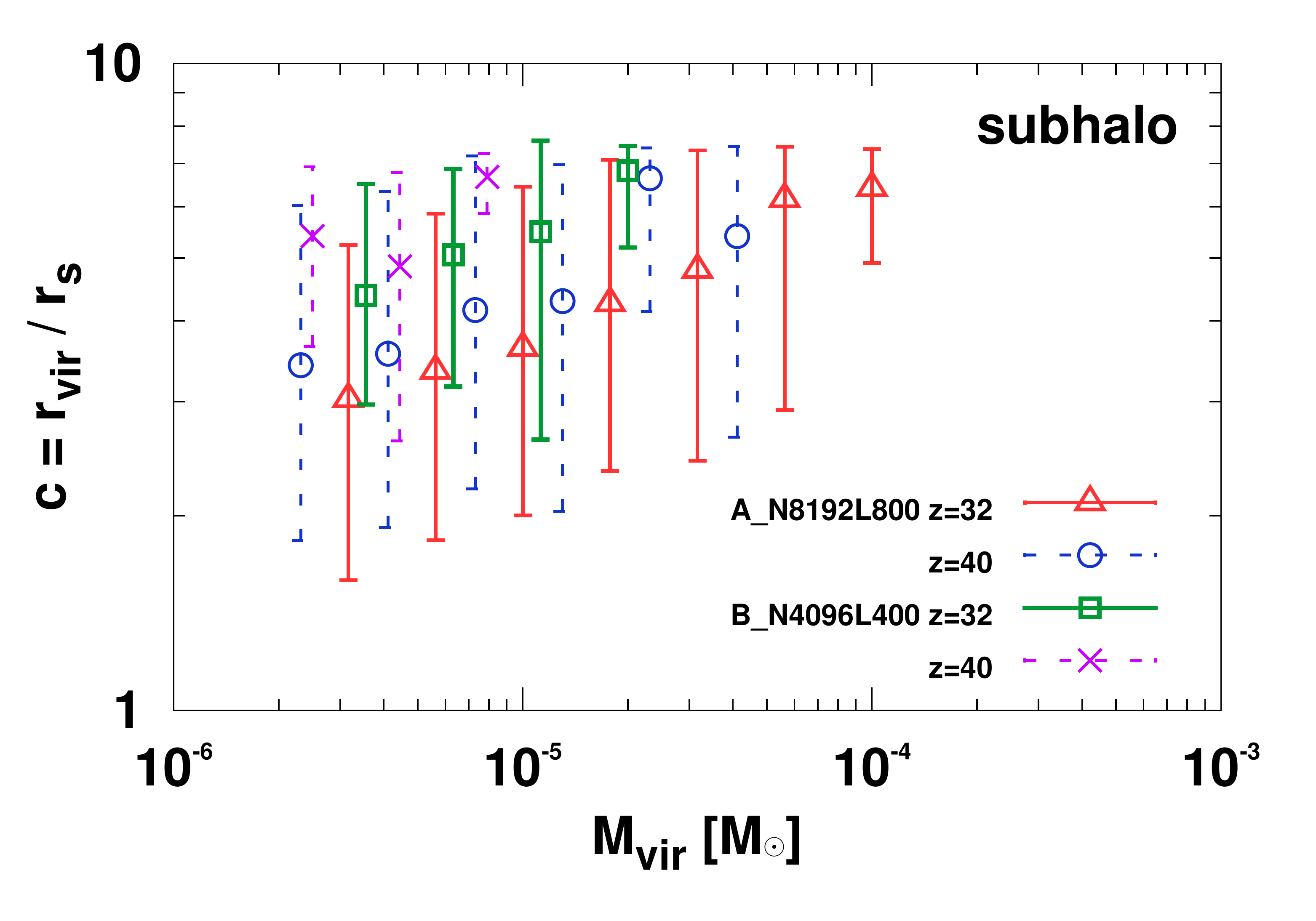}
\caption{ 
Evolution of the slopes (left) and concentrations (right) of 
subhaloes as a function of the virial mass \mvir. 
  Triangles, circles, squares, and crosses 
  are the median value in each mass bin.  Whiskers give
  the first and third quantiles. 
}
\label{fig:profile_evo}
\end{figure*}

These results are obtained in field haloes, not subhaloes.  The
structures of subhaloes should be different from field haloes because of
tidal stripping.  Cosmological simulations for more massive haloes have shown
that subhaloes are more concentrated than field haloes
\citep[e.g.,][]{Ghigna2000,Bullock2001,Moline2017}.  
To estimate the annihilation signal more robustly, 
quantifying the structures of
subhalo near the free streaming scale is necessary.

I14 has carefully performed resolution studies and have
conservatively concluded that the inner density profile at 2\% of
their virial radius is well resolved for the haloes with masses more
massive than $2 \times 10^{-6}$\Msun\ under the resolution of \Af\ and
\Bf\ simulations.  Below this radius, numerical two-body relaxation
becomes serious.  The mass resolution and softening parameter of
\Afs\ and \Bfs\ simulations are comparable to those of the
\Af\ simulation, which is used in I14.  Thus, we hereafter use the
same criterion to quantify the density profiles of haloes and
subhaloes in this study.

We calculated the spherically averaged radial density
profile of each subhalo over the range $0.02 \le r/r_{\rm vir} \le
1.0$, splitted into 32 logarithmically equal bins.  To exclude
dynamically unstable haloes and subhaloes, we add another criterion
$K/|W| < 1.0$, where $K$ and $W$ are internal kinetic and potential
energies of each halo.  We use Equation~(\ref{eq:doublepower}) to
quantify the density profiles of haloes and subhaloes.
The host halo mass range is between $\sim 5 \times 10^{-6}$ 
and $\sim 3 \times 10^{-2}$\Msun. 

Figure~\ref{fig:profile_a} and \ref{fig:profile_c} visualize the
median and scatter of the slope $\alpha$ and the concentration as a
function of the field halo and subhalo mass at $z=32$.  Only mass bins
containing more than 20 haloes (subhaloes) are shown.  The slopes of
field haloes show the stark difference between with and without the
cutoff. The slope $\alpha$ is almost constant ($\alpha \sim 1.1$) in
the no cutoff simulation (\Bfs), but is substantially steeper
and has mass dependence in the cutoff simulation (\Afs).
These results are consistent with I14.

The central slopes are considerably shallower in subhaloes than field haloes 
for both simulations with and without the cutoff. For the cutoff 
simulation (\Afs), the mass dependence is more prominent in subhaloes 
than in field haloes. For the no cutoff simulation (\Bfs), the mass 
dependence emerges in subhaloes differently from field haloes. These 
difference should result in the effect of tidal stripping from host 
haloes. 

For field haloes, the concentrations in both simulations are almost
constant regardless of the halo mass over the range shown in
Figure~\ref{fig:profile_c}.  The median concentration in the cutoff
model is about 1.5, increasing to 2.0 without the cutoff.  These
values are slightly larger than those observed in I14, possibly
because of the difference of the criterion to select haloes.  The
dynamical condition $K/|W| < 1.0$ is imposed in this study but is not
in I14.  Figure~\ref{fig:profile_c} indicates clearly that the
concentrations are significantly larger in subhaloes than haloes and
depend on the subhalo mass because of the tidal stripping.  This
picture is qualitatively consistent with what we see in more massive
haloes \citep[e.g.,][]{Ghigna2000,Bullock2001,Moline2017}.

Comparing with field haloes, the slope and concentration of subhaloes contain
larger scatters probably because of the tidal stripping. How host haloes 
perturb the structure of subhaloes should depend on when they are 
accreted on and their orbit.  The variation of these parameters should 
increase the scatter of the slope and concentration. The detail of the 
scatter is beyond the scope of this paper and will be addressed in 
future works.

Figure~\ref{fig:profile_evo} depicts the redshift evolution of the slope
and concentration of subhaloes at $z=32$ and 40.  There are no large
differences on these properties between $z=32$ and 40 for both
simulations with and without the cutoff.  The power law functions that
give best fits with the relation between mass and the shape parameter $\alpha$ of field
haloes of $10^{-6}\sim10^{-3}{\rm M_{\odot}}$ 
and subhaloes of $10^{-6}\sim10^{-4}{\rm M_{\odot}}$ are
\begin{eqnarray}
\alpha_{\rm halo}(M_{\rm vir}) &=& 
-0.123 \log_{10}( M_{\rm vir} / 10^{-6}{\rm M_{\odot})} + 1.461, \label{eq:fita}  \\
\alpha_{\rm subhalo}(M_{\rm vir}) &=& 
-0.296 \log_{10}( M_{\rm vir} / 10^{-6}{\rm M_{\odot}}) + 1.473. \label{eq:fita2}
\end{eqnarray}
The relation for field haloes is taken from I14 
and is consistent with the simulations in this work.

\subsection{Extrapolating the mass-concentration relation to other 
redshifts}\label{sec:extrapolate}

\begin{figure*}
\centering 
\includegraphics[width=8.5cm]{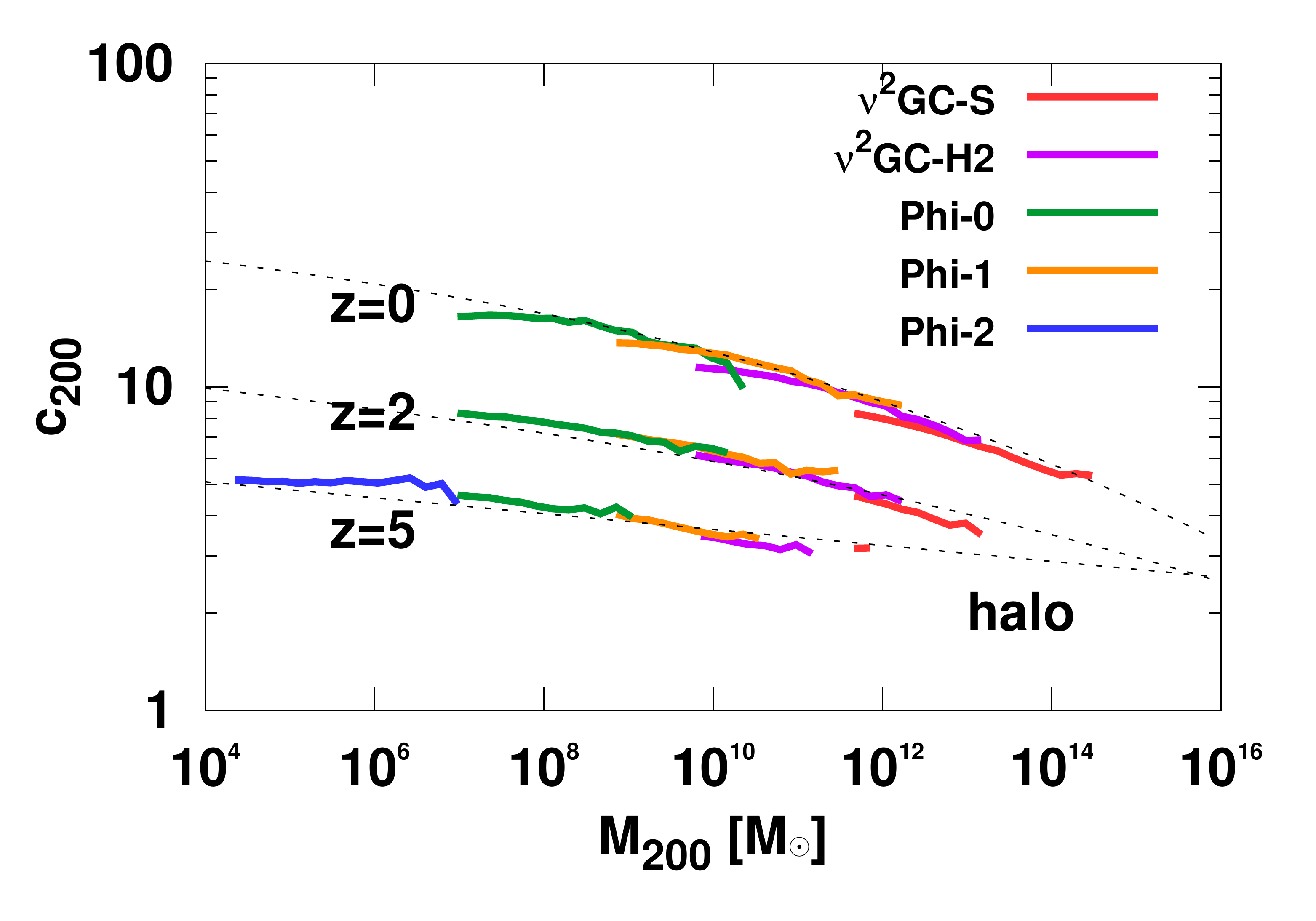}
\includegraphics[width=8.5cm]{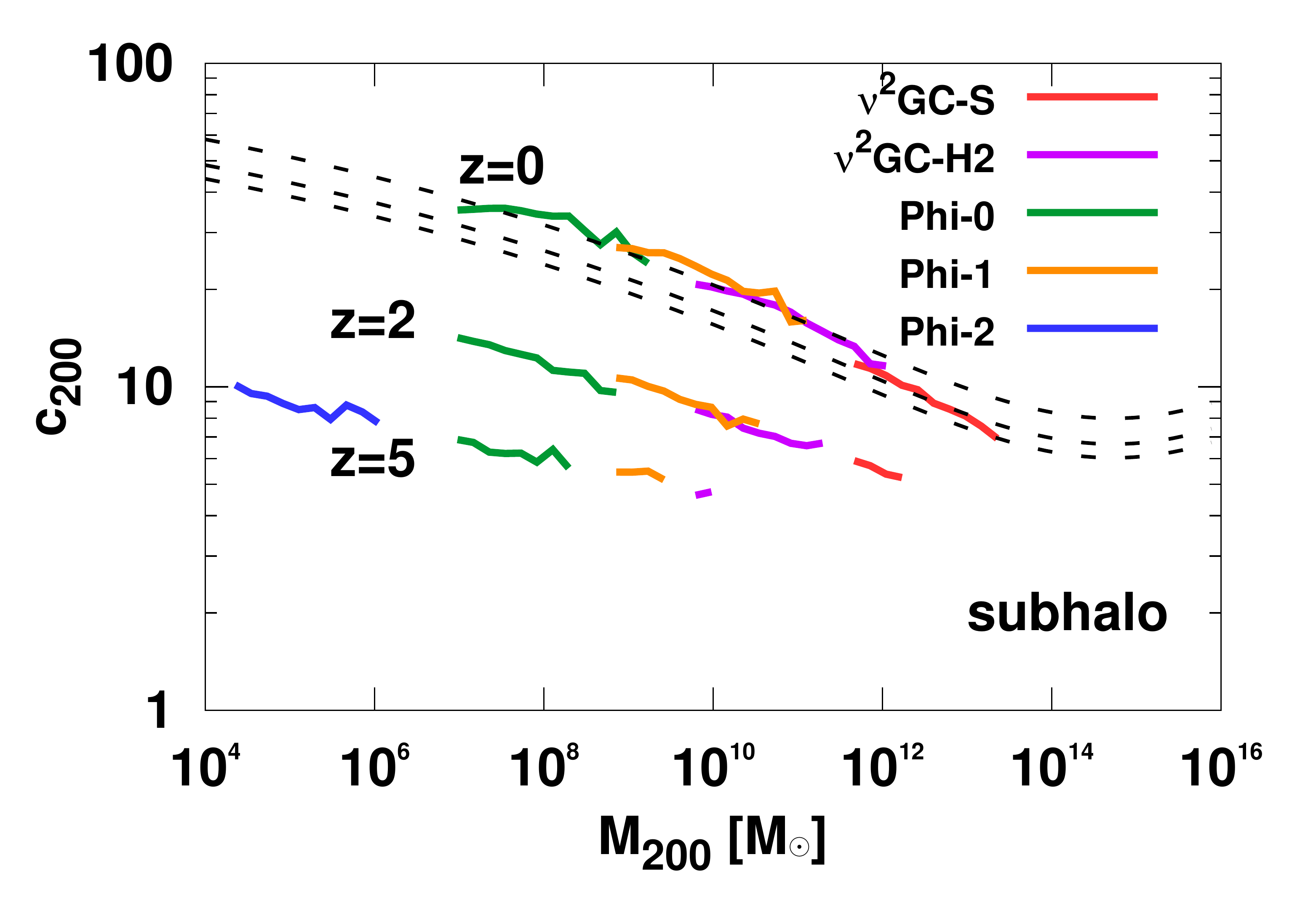}
\caption{ 
Concentrations $c_{200}$ of field haloes (left) and subhaloes (right) as a
function of the halo mass $M_{\rm 200}$.  Upper, middle, and lower
solid curves are results at $z=0, 2$, and 5, respectively.  Three
thin dashed curves in the left panel are fitting functions proposed by
\citet{Correa2015b}.  Three thick dashed curves in the right panel show the
model of \citep{Moline2017} with $x_{\rm sub}=0.1, 0.3$, and 0.5 (upper to
bottom), respectively.
}
\label{fig:m200-c200}
\end{figure*}

\begin{figure*}
\centering 
\includegraphics[width=8.8cm]{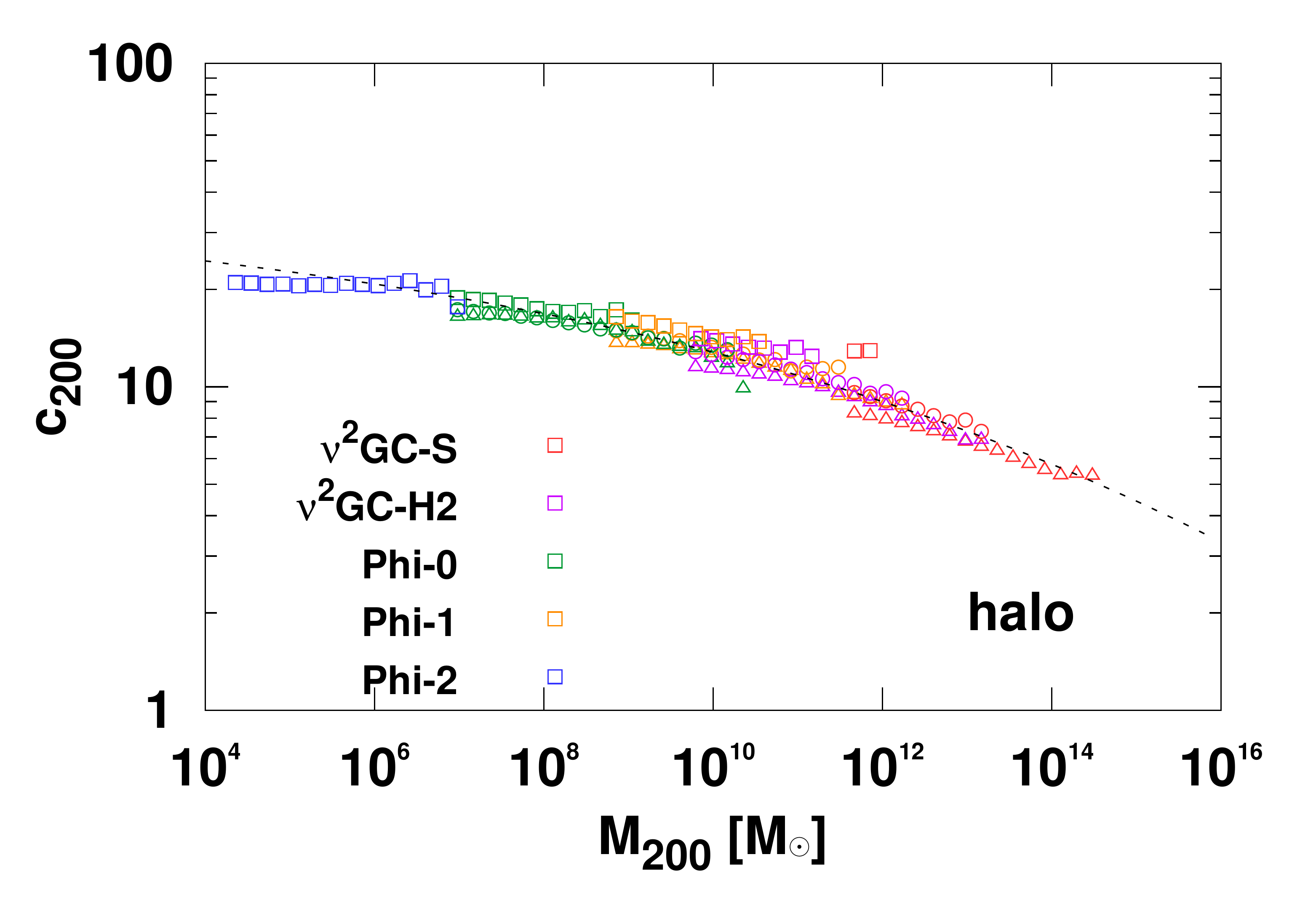}
\includegraphics[width=8.8cm]{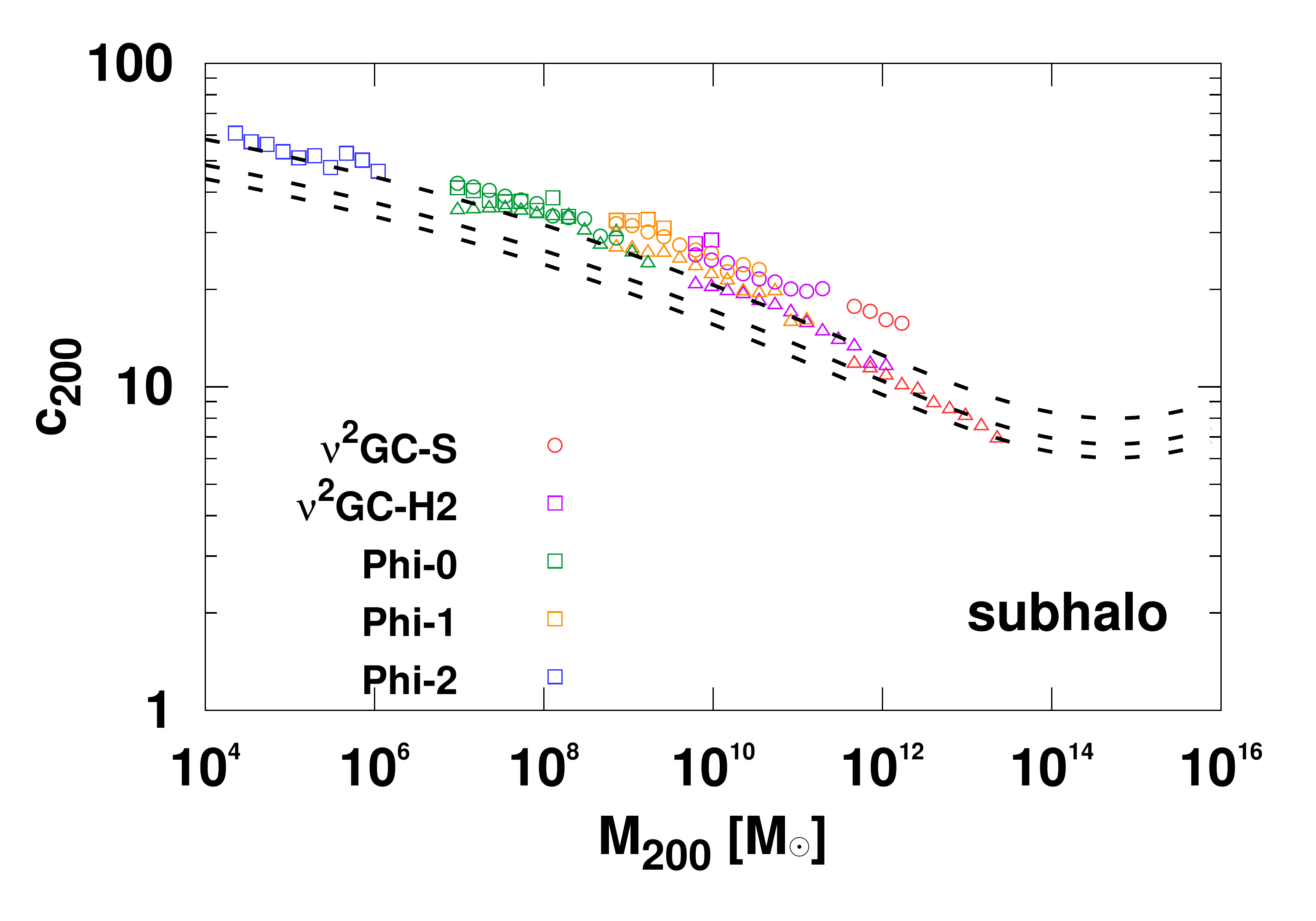}
\caption{ 
Scaled concentrations of field haloes (left) and subhaloes (right) to $z=0$
as a function of the halo mass $M_{200}$. 
For field haloes, the concentrations are scaled by multiplying $[H(z)/H(0)]^{2/3}$. 
For subhaloes, the scaling $(1+z)$ is used. 
Triangles, circles, and squares show the results at $z=0, 2, 5$, respectively.
Thin dashed curves in the left panel are fitting functions proposed by
\citet{Correa2015b}.  Three thick dashed curves in the right panel show the
model of \citet{Moline2017} with $x_{\rm sub}=0.1, 0.3$, and 0.5 (upper to
bottom), respectively.
}
\label{fig:m200-c200_scale}
\end{figure*}

\begin{figure*}
\centering 
\includegraphics[width=8.8cm]{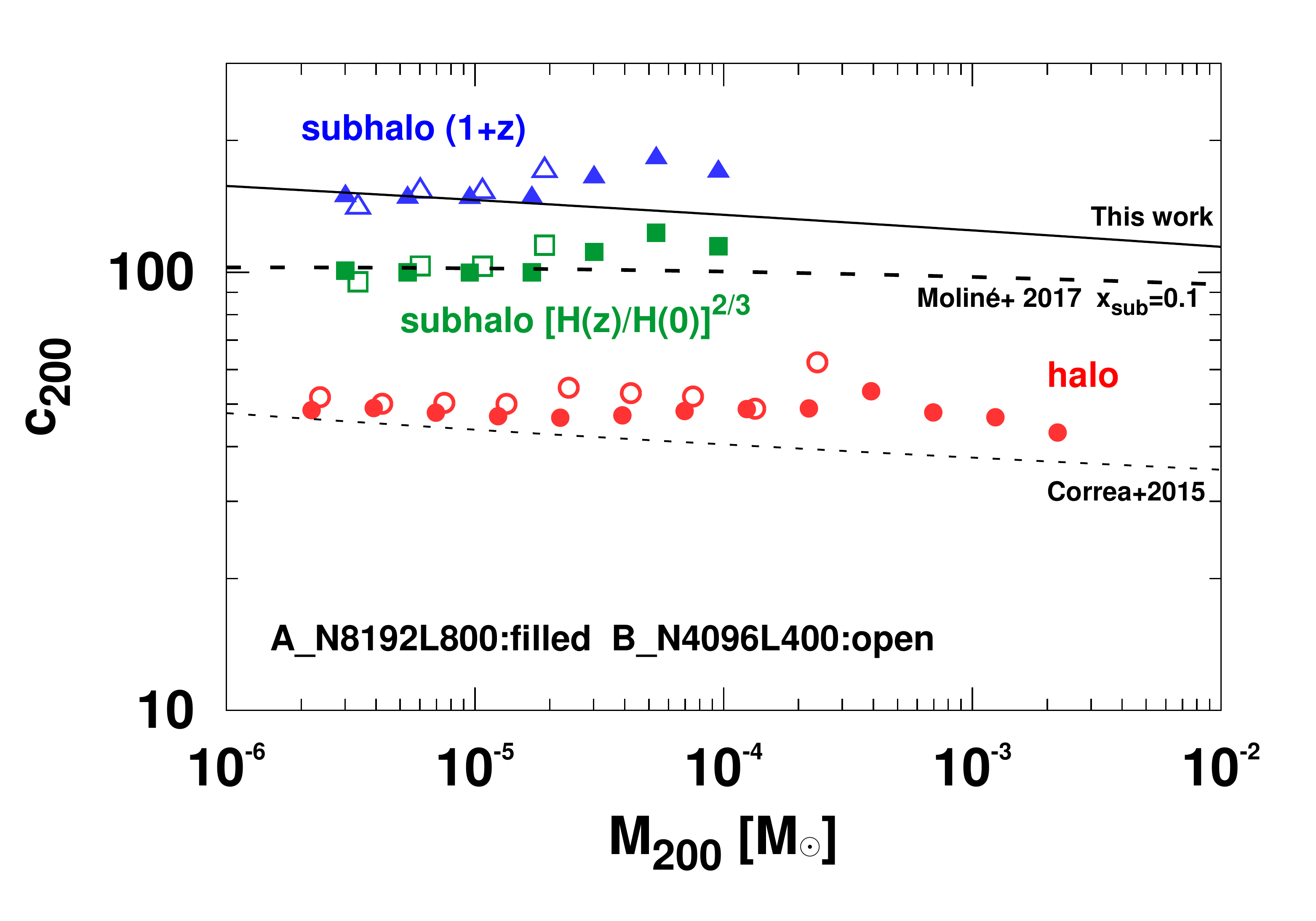}
\includegraphics[width=8.8cm]{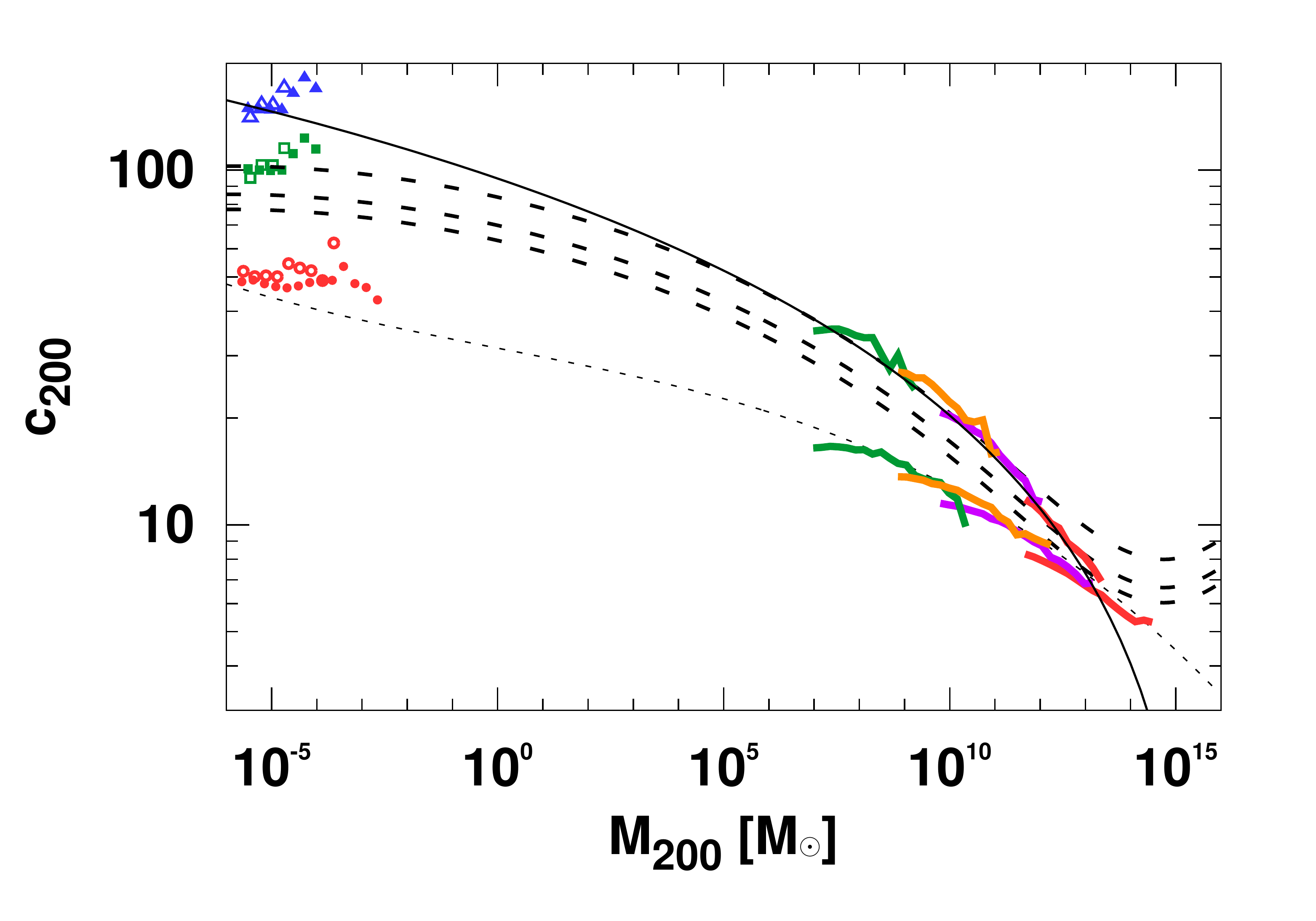}
\caption{ 
Converted concentrations of the density profiles of haloes and subhaloes 
near the free streaming scale versus the halo mass.
As described in \S~\ref{sec:profile}, 
the concentrations are converted to counterparts of the NFW profile at $z=0$. 
Thin and thick dashed curves are fitting functions proposed by
\citet{Correa2015b} and \citet{Moline2017} with $x_{\rm sub}=0.1$, respectively. 
Solid curve is the fitting function proposed in this work 
(Equation~(\ref{eq:m-c}) in the text). 
  }
\label{fig:profile_nfw}
\end{figure*}

To estimate the contributions of subhaloes near the free streaming scale 
to gamma-ray annihilation signals, extrapolating the density profiles 
to $z=0$ is necessary.  
Under the assumption that the shape parameter is unchanged with the redshift,
the commonly used extrapolation of the 
concentration is the multiplier $(1+z)$ \citep[e.g.,][]{Bullock2001} and 
$[H(z)/H(0)]^{2/3}$\citep{Maccio2008, Pilipenko2017}, where $H(z)$ is 
the Hubble constant.  However, the validity of these scaling for 
subhaloes and less massive field haloes are not fully understood.

We first test the evolution of the concentration of subhaloes and less 
massive haloes from five cosmological $N$-body simulations as described 
in Table~\ref{tab2}.  We selected all haloes and subhaloes that contain 
more than 1,000 particles.  
The mass-concentration relation for field haloes
has been refined until today
\citep[e.g.,][]{Prada2012, Correa2015b, Klypin2016, Okoli2016, Pilipenko2017, Child2018, Diemer2019} 
and that for less 
massive field haloes has been studied down to $\sim 10^7$\Msun\ to date 
\citep[e.g.,][]{Pilipenko2017}. It is also studied for near the free 
streaming scale \citep[$\le \sim 10^{-3}$\Msun, ][]{Ishiyama2014}. However, 
the relation between $10^{-3}$ and $10^{7}$\Msun\ is not understood well. 
Our simulations enable us to test first it down to $\sim 10^4$\Msun.

We use $M_{\rm 200}$ and
$c_{\rm 200}$ instead of $M_{\rm vir}$ and $c_{\rm vir}=r_{\rm
  vir}/r_{\rm s}$ to compare easily with relevant studies
\citep[e.g.,][]{Correa2015b, Moline2017}.  Here,
$M_{\rm 200}$ is the enclosed mass within $r_{\rm 200}$ in which the
spherical overdensity is 200 times the critical density in the
Universe.  Normally, $c_{\rm 200}$ is defined by $c_{\rm
  200}=r_{200}/r_{\rm s}$, however, we calculate it using the maximum
circular velocity and the circular velocity at $r_{200}$ with the
assumption of the NFW profile, according to the method described in
\citet{Klypin2011, Prada2012, Pilipenko2017}.

Figure~\ref{fig:m200-c200} shows the mass-concentration relations of 
field haloes and subhaloes as a function of $M_{\rm 200}$ from $\sim 10^4$ 
to $\sim 10^{15} {\rm M_{\odot}}$ at $z=0, 2$ and 5.
The fitting functions proposed by \citet{Correa2015b}
well describe the concentrations of field haloes.
On the other hand, 
concentrations are significantly larger in subhaloes than field haloes, 
and the slope of mass-concentration relation is steeper in subhaloes.  
The subhalo concentration at $z=0$ agrees well with the fitting function 
proposed by \citet{Moline2017}, 
which also includes the dependence of the distance 
between the centre of the host halo and the subhalo
, $x_{\rm sub}$. 

The multiplication $[H(z)/H(0)]^{2/3}$ with the concentrations of field haloes
works well, as shown in Figure~\ref{fig:m200-c200_scale}.  On the other
hand, interestingly, the multiplication $(1+z)$ well matches for
subhaloes over the broad mass range.  We do not pursue the physical
origin of this scaling difference here because it is beyond the scope
of this paper.  We evaluate the difference between both scaling on the
annihilation signal in \S~\ref{sec:discussion}.

Finally, we convert the mass-concentration relation of haloes and 
subhaloes near the free streaming scale from $z=32$ to 0.  However, our 
simulations reveal that the density profiles in the cutoff simulation 
significantly deviate from the NFW profile and have the dependence on 
the halo mass. Therefore, our results should not be directly comparable with 
the mass-concentration relation proposed by other studies 
assuming the universal NFW profile.

To perform an indirect comparison, we converted the concentrations to 
that of the NFW profile by a method used in other literature 
\citep[e.g.,][]{Ricotti2003, Anderhalden2013,
  Ishiyama2014}.  The concentration in the profile of 
Equation~(\ref{eq:doublepower}) can be converted to the equivalent NFW
concentration by multiplying $1.0/(2.0-\alpha)$.  The extrapolation
from $z=32$ to 0 is tested by multiplying $[H(z)/H(0)]^{2/3}$ for haloes
and both $[H(z)/H(0)]^{2/3}$ and $(1+z)$ for subhaloes.

Figure~\ref{fig:profile_nfw} plots the converted mass-concentration relation
of haloes and subhaloes for the \Afs\ and \Bfs\ simulations against halo
mass.  The halo concentration shows excellent agreement with
\citet{Correa2015b}, which is calibrated by more massive haloes (galaxy
to cluster scale) and lower redshifts, although there is the small
systematic upper shift.
The redshift scaling of \citet{Correa2015b} is slower than 
both $[H(z)/H(0)]^{2/3}$ and $(1+z)$, 
explaining this difference to some degree. 
The halo concentration is slightly smaller 
those found in earlier simulations \citep{Ishiyama2014} 
because of the usage of the scaling $(1+z)$
in the literature, which 
gives the higher concentrations than the scaling $[H(z)/H(0)]^{2/3}$.

The subhalo concentration with the scaling
$[H(z)/H(0)]^{2/3}$ agrees well with the model of \citet{Moline2017}
using $x_{\rm sub}=0.1$.
The subhalo concentration with the scaling
$(1+z)$ gives the largest concentration. 
We suggest a simple fitting function of this subhalo concentration-mass relation 
for the smallest to the largest resolved scale ($10^{-6}\sim10^{12}{\rm M_{\odot}}$)
as,
\begin{eqnarray}
c_{200} = \sum^3_{i=0} c_i \times \left[
  \ln\left(\frac{m_{200}}{\rm M_{\odot}}\right) \right]^i, 
\label{eq:m-c}
\end{eqnarray}
where, 
$c_i = [94.6609, -4.1160, 3.3747\times10^{-2}, 2.0932\times10^{-4}]$. 
This simple functional form is the same used in 
\citet{Lavalle2008, Sanchez-Conde2014}. 
These parameterizations agree well with the subhalo 
mass-concentration relation as shown in Figure~\ref{fig:profile_nfw} 
and also match 
it in more massive subhaloes from $\sim 10^4$ to
$\sim 10^{12}$\Msun.

\section{Discussions}\label{sec:discussion}

In this section, we evaluate the impact of 
the subhalo mass function and the subhalo density profile
obtained in this paper on the annihilation boost factor.
All subhaloes contribute the annihilation luminosity of a host halo.
The boost factor $B(M)$ is given by \citep[e.g.,][]{Strigari2007, Moline2017}
\begin{eqnarray}
B(M) &=& \frac{3}{L_{\rm host}(M)} \int^M_{M_{\rm min}}\frac{dn}{dm} \,dm \int^1_0 \, dx_{\rm sub} \times \nonumber \\ 
&& [1+B_{\rm sub}(m)] \, L_{\rm sub}(m,x_{\rm sub}) \, x^2_{\rm sub} 
\label{eq:bf}
\end{eqnarray}
where, $L_{\rm host}(M)$ and $L_{\rm sub}(M,x_{\rm sub})$ is the annihilation luminosity of a halo and 
subhalo of mass $M$ with a smooth distribution (without subhaloes), 
$dn/dm$ is the subhalo mass function, 
and $x_{\rm sub}=r/r_{200}$.
Here, $r$ is the the distance from the centre of the host halo, 
and we incorporated the dependence of the concentration 
on $r$ by the way described in (iv) below.
We computed the annihilation luminosity of each mass halo 
by performing numerical integration 
in volume
of the square of Equation~(\ref{eq:doublepower}), 
from $10^{-5} \rm pc$ to tidal radii of subhaloes 
(See (v) below). 

Based on the results of previous sections,
we use the following model to estimate the annihilation boost factor 
at $z=0$.
\begin{enumerate}
\item
Density profiles of host haloes: 
We assume the NFW profile for host haloes and 
the mass-concentration relation proposed by \citet{Correa2015b}.
\item 
Subhalo mass function: 
As a fiducial model, we use a fitting formula proposed by \citet{Ando2019}, 
which is based on a successful semianalytic model of subhaloes \citep{Hiroshima2018}. 
For comparison, we also adopt the subhalo mass function, 
$dn/dm = 0.012M^{-1} (m/M)^{-2.0}$
\citep[e.g.][]{Sanchez-Conde2014, Moline2017}, 
which however overpredicts the subhalo abundance \citep{Hiroshima2018}. 
To incorporate the effect of cutoff, 
we multiply the subhalo mass functions by the 
correction functions given by Equation~(\ref{eq:fcor}).
We use $M_1=1.3 \times 10^{-6}$ and $M_2=1.0 \times 10^{-7}$ 
(correction function 2) by default. 
however we also test $M_2=1.0 \times 10^{-8}$ (correction function 1).
The smallest limit of the integral (\ref{eq:bf}) is $M_{\rm  min}=10^{-6}$\Msun.
\item 
Density profiles of subhaloes: 
The slope $\alpha$ in subhalo density profiles are described by 
Equation~(\ref{eq:fita2}). When they give values smaller than one, 
$\alpha$ is forced to be one under the assumption that the profile is like the NFW profile. 
Although $\alpha$ is less than one for the most massive two bins as 
seen in Figure \ref{fig:profile_c}, 
we force to be one to smoothly connect to the profile of more massive subhalos, 
which is reasonably well fitted by the NFW profile \citep{Springel2008}.
\item 
The mass-concentration relation of subhaloes: 
If $\alpha>1$,
the concentration is converted to that of the double power law function
employing the opposite way described in \S \ref{sec:extrapolate}.
Extrapolation to z=0 is done by multiplying $(1+z)$. 
We assume the average concentration as a function of subhalo mass 
is described by Equation~(\ref{eq:m-c}), and 
modify it to incorporate the dependence of the concentration 
on the distance from the centre of the host halo $r$, as 
$0.95 c_{200}(m_{200}) [1.0 - 0.54\log_{10}(x_{\rm sub})]$, 
where $c_{200}(m_{200})$ is the same with Equation~(\ref{eq:m-c}) and 
$x_{\rm sub}=r/r_{200}$.
Dependence on the distance is following \citet{Moline2017}, and 
the normalization is determined with the assumption 
that $0.81r_{200}$ is the average position of subhaloes \citep{Springel2008}.
As described in
\S \ref{sec:extrapolate}, the extrapolation of the
mass-concentration relation has considerable uncertainty.  
Thus, we also
compute the boost factor using the fitting function proposed by
\citet{Moline2017}, which gives quantitatively good agreement with our
simulation results when the scaling $[H(z)/H(0)]^{2/3}$ is used.
We integrated $x_{\rm sub}$ from the centre to 1 ($r=r_{200}$) of the host halo
assuming subhaloes distribute uniformly. 
We call the former ``high concentration'' and 
the latter ``low concentration'' model.
\item Tidal radius: 
We calculate the tidal radii of subhaloes 
by following \citet{Moline2017}.
\end{enumerate}

\begin{figure}
\centering 
\includegraphics[width=8.8cm]{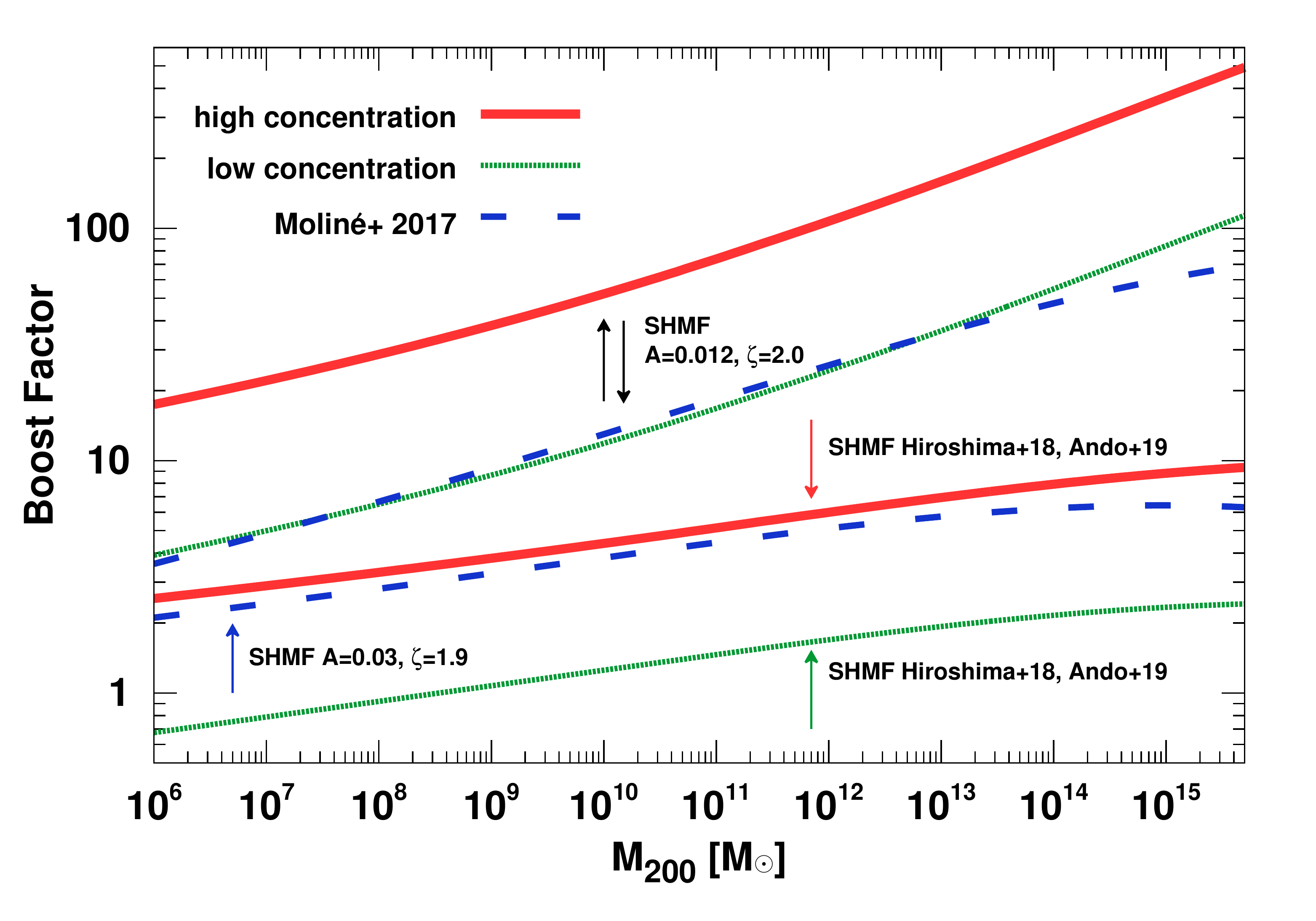} 
\caption{ 
Annihilation boost factor as a function of the host halo mass.
Two different mass-concentration relations for subhaloes are used. 
Upper solid and dotted 
curves are our model results with the subhalo mass function 
(SHMF) $dn/dm = AM^{-1} (m/M)^{-\zeta}$ ($A=0.012$ and $\zeta=2.0$). 
Bottom solid and dotted
curves are results with the subhalo mass function proposed by 
\citet{Ando2019}, which is based on a successful 
semianalytic model of subhaloes \citep{Hiroshima2018}. 
Dashed curves are from \citet{Moline2017} with two different
subhalo mass functions. Upper and lower ones use $A=0.012, \zeta=2.0$ 
and $A=0.03, \zeta=1.9$, respectively.
}
\label{fig:bf}
\end{figure}

Figure~\ref{fig:bf} shows the annihilation boost factor versus the host halo mass.
For the fiducial subhalo mass function \citep{Ando2019},
the boost factors of a Milky-Way sized halo ($M \sim 2.0 \times 10^{12} M_{\odot}$) 
are $\sim 6.23$ and $1.75$ for 
models of ``high'' and ``low concentration'', respectively. 
Those with another subhalo mass function are $\sim 118$ and $27$, respectively. 
In the latter case, 
our models raise the boost factor substantially, 
by a factor of four
compared with that using the mass-concentration relation of field haloes, 
which gives the boost factors $\sim 29$ \citep{Ishiyama2014}. 
However, such considerable boost might be unrealistic because 
the fiducial subhalo mass function is favoured \citep{Hiroshima2018, Ando2019}. 

When we apply the other correction function for the subhalo mass
function (correction function 1), the resulting boost factor of a
Milky-Way sized halo is nearly the same with the other correction
function.
Without the correction function of the subhalo mass function, 
the boost factor of a Milky-Way sized halo becomes 6.65 for the 
``high concentration'' model. 
On the other hand, when we force the NFW profile for all subhaloes, 
the resulting boost factor is 5.96. 
Thus, the suppression of the subhalo abundance 
near the free streaming scale reduces the boost by $\sim6\%$ 
and the steeper cusp increase it by $\sim5\%$, 
indicating that both effects compensate each other. 
This explains the reason that the boost factors of our model and 
\citet{Moline2017} agree well with each other when the 
same subhalo mass function and the mass-concentration relation 
for subhaloes are used. 
The deviation at high mass end reflects the difference of 
adopted mass-concentration relation for host haloes. 
The relation of \citet{Sanchez-Conde2014} adopted in \citet{Moline2017}
gives higher concentrations at high mass end than that of \citet{Correa2015b} 
adopted in this work. 
Thus, resulting boost factors are smaller in \citet{Moline2017}.

There is a good agreement between 
the model of high concentration with the fiducial subhalo mass function 
and the model of \citet{Moline2017} with $dn/dm = 0.03M^{-1} (m/M)^{-1.9}$. 
The latter subhalo mass function gives a larger number of subhalos 
than the fiducial one, not so much as that with the slope $-2$. 
This is a reason that both boost factors agree well,
whereas the concentration in the former is substantially high. 

In these calculations, we ignore the contribution of subhaloes below $10^{-6} M_{\odot}$. 
The density profile of haloes and subhaloes below the cutoff scale is not 
understood well, and 
the assumption for it in our model might be invalid. 
We will address this subject in a future paper. However, our results 
suggest that the existence of the cutoff on the subhalo mass function 
could obscure this uncertainty and allow us to estimate the annihilation 
signal robustly if the mass-concentration relation is given correctly.

\section{Summary}\label{sec:summary}

We have studied the abundance and structure of subhaloes near the free 
streaming scale using a suite of unprecedentedly large 
cosmological $N$-body simulations.  We used two different models of 
initial matter power spectra with and without the cutoff, which is 
resulted from the free streaming damping of WIMP dark matter 
particle.  We have investigated the effect of cutoff on the subhalo mass 
function and the density profile of subhalo. Our primary results are 
summarized below.

\begin{enumerate}
\item 
For the range of host halo mass between $10^{-4}$ and $10^{-2}$\Msun, the
subhalo mass functions in the cutoff simulation agree with those in
the no cutoff simulation for masses more massive than $\sim 5.0
\times 10^{-6}$\Msun.  For the less massive subhaloes, the slopes of
mass functions $\zeta$ are gradually decreasing with decreasing subhalo mass
differently from the no cutoff simulation, 
and are becoming flat at
around $\sim 10^{-6}$\Msun, which corresponds to the cutoff scale.
The slopes $\zeta$ are becoming steeper again from $\sim
10^{-7}$\Msun\ due to artificial fragmentation as seen in warm dark
matter simulations.  The ratio between the subhalo mass function in
the cutoff and no cutoff simulation is well fitted by the correction
function described in Equation~(\ref{eq:fcor}),
regardless of the host halo mass and the redshift.
\item 
In subhaloes, the central slopes are considerably shallower
than in field haloes for both simulations with and without the
cutoff, but are still steeper than that of the NFW profile. 
The shape parameter $\alpha$ is given by
$-0.296 \log_{10}( M_{\rm vir} / 10^{-6}{\rm M_{\odot}}) + 1.473$.
\item 
The concentrations are significantly larger in subhaloes than haloes and
depend on the subhalo mass because of the tidal stripping.  This
picture is qualitatively consistent with what we see in more massive
haloes \citep[e.g.,][]{Ghigna2000,Bullock2001,Moline2017}.
\item
We compare two 
methods to extrapolate the mass-concentration relation of haloes and 
subhaloes to z=0 and provide a new simple fitting function for subhaloes, 
based on a suite of large cosmological $N$-body simulations.
Finally, we estimate the annihilation boost factor of a Milky-Way sized halo
to be between 1.8 and 6.2. 
\end{enumerate}

\section*{Acknowledgments}
We thank the anonymous referee for his/her valuable comments.
We thank Jorge Pe$\tilde{\rm n}$arrubia, Daisuke Nagai, Sheridan
Green, and Nagisa Hiroshima for helpful discussions.  Numerical
computations were partially carried out on the K computer at the RIKEN
Advanced Institute for Computational Science (Proposal numbers
hp150226, hp160212, hp170231, hp180180, hp190161), Aterui and Aterui II
supercomputer at Center for Computational Astrophysics, CfCA, of
National Astronomical Observatory of Japan.  This work has been
supported by MEXT as ``Priority Issue on Post-K computer''
(Elucidation of the Fundamental Laws and Evolution of the Universe)
and JICFuS. We thank the support by MEXT/JSPS KAKENHI Grant Number
JP15H01030 (TI), JP17H04828 (TI), JP17H01101 (TI), JP18H04337 (TI), 
JP17H04836 (SA), JP18H04340 (SA), and JP18H04578 (SA).

\bibliographystyle{mnras}

\bsp	
\label{lastpage}
\end{document}